\def\year{2018}\relax
\documentclass[letterpaper]{article} %DO NOT CHANGE THIS
\usepackage{aaai18}  %Required
\usepackage{times}  %Required
\usepackage{helvet}  %Required
\usepackage{courier}  %Required
\usepackage{url}  %Required
\usepackage{graphicx}  %Required
\usepackage{soul}
\usepackage{amsmath,amssymb}
\usepackage{xr}
\usepackage[labelformat=simple]{caption}
\usepackage{subcaption}
\usepackage{multicol}

\title{Using Simpson's Paradox to Discover Interesting Patterns in Behavioral Data}
\author{Nazanin Alipourfard, Peter G. Fennell	 \and Kristina Lerman\\
USC Information Sciences Institute\\
4676 Admiralty Way\\
Marina del Rey, California 90292\\
}
\begin{document}
\maketitle

\graphicspath{ {figures/} }

\maketitle
\begin{abstract}
We describe a data-driven discovery method that leverages Simpson's paradox to uncover interesting patterns in behavioral data. Our method systematically disaggregates data to identify subgroups within a population whose behavior deviates significantly from the rest of the population. Given an outcome of interest and a set of covariates, the method follows three steps. First, it disaggregates data into subgroups, by conditioning on a particular covariate, so as minimize the variation of the outcome within the subgroups. Next, it models the outcome as a linear function of another covariate, both in the subgroups and in the aggregate data. Finally, it compares trends
to identify disaggregations that produce subgroups with different behaviors from the aggregate.
We illustrate the method by applying it to three real-world behavioral datasets, including Q\&A site Stack Exchange and online learning platforms Khan Academy and Duolingo.
\end{abstract}

% ______________________________________________________________________
%                                                 INTRODUCTION
% ______________________________________________________________________
\section{Introduction}
Digital traces of activity have exposed human behavior to quantitative analysis~\cite{Lazer09,mcfarland2016sociology}. Data mining algorithms have explored behavioral data to test social psychology and decision theories~\cite{kleinberg2017human,bond201261} and obtain new insights into factors affecting online behavior.
Yet, behavioral data analysis is still largely a trial-and-error process, driven by ad-hoc methods rather than principled solutions. Compounding the difficulty are the multi-faceted challenges presented by behavioral data: it is massive, multi-dimensional, noisy, sparse (few observations per individual), heterogeneous (composed of differently behaving individuals), and highly unbalanced (very few observations of the outcome of interest exist). As a result, given a new behavioral data set, it is often unclear where to start or how to even go about identifying interesting phenomena in data. To explore a new data set, a researcher may do exploratory analysis, for example,  plot the distributions of features of interest or perform principal component analysis, but beyond this, lack of clear guidelines for analytic practice make quantitative exploration of large-scale behavioral data more of an art than a science.

The current paper takes a step towards solving this problem by automating discovery from behavioral data. We propose a method that systematically uncovers surprising patterns in data by identifying subgroups within the population whose behaviors are substantially different from the rest of the population.
Our method leverages Simpson's paradox~\cite{Blyth1972,Norton2015simpson}, a phenomenon wherein an association or a trend observed in the data at the level of the entire population disappears or even reverses when data is disaggregated by its underlying subgroups. %However, identifying the subgroups that differ in their behavior is challenging, because their composition within the population may change as a function of some covariate in the data~\cite{Norton2015simpson,Alipourfard2018}.
The goal of our method is to identify a covariate, such that conditioning data on the covariate significantly changes the association between the outcome and another covariate (acting as an independent variable).
To address this challenge, we introduce \emph{Simpson's Disaggregation}, a method that decomposes the population into subgroups and compares behavioral trends within subgroups to find surprising patterns. First, our method identifies potential subgroups by disaggregating the data into bins that minimize the variation of the outcome of interest. It then uses a linear model to represent behavioral trends within subgroups, as well as in aggregate data, and looks for trend reversal.
Finally, it uses statistical methods to assess the significance of trends in both aggregated and disaggregated data, and compares disaggregations based on their explanatory power.

We apply the fully automatic method to several real-world behavioral data sets that include Q\&A site Stack Exchange, online learning platforms Khan Academy and Duolingo. These data sets are all highly heterogeneous. After disaggregating the data, we find that the trends describing the response of the outcome to various covariates within the subgroups can be very different from the population-level response. We show that disaggregations lead to models that better explain the data.   We uncover common patterns across data sets about the effects of skill and experience on user performance and suggest further lines of inquiry into behaviors on these platforms.

By dissecting the data into more homogeneous subgroups, our method can uncover surprising subgroups that behave differently from the rest of the population. Such patterns are a sign that strong individual differences exist within the population, differences that must be accounted for in analysis. Thus, the method gives a researcher a powerful tool for automatically identifying behavioral patterns meriting deeper study.

The rest of the paper is organized as follows. First, we review Simpson's paradox and methods used to identify it. Next, we describe a method that automatically partitions the data into more homogenous subgroups and identifies surprising trends. In the Results section, we apply the method to real-world behavioral data and show that our method is able to identify interesting phenomena in data.

% ______________________________________________________________________
%                                                 RELATED
% ______________________________________________________________________
\section{Background and Related Work}
\label{sec:related}
Simpson's paradox~\cite{Blyth1972,Norton2015simpson} often confounds analysis of heterogeneous data, especially social and behavioral data. According to the paradox, an association observed in data that has been aggregated over the entire population may be different from, and even opposite to, associations existing in the subgroups comprising the population. Failure to take this paradox into account can distort conclusions of analysis. Arguably the most famous example of Simpson's paradox arose from a lawsuit alleging gender discrimination at UC Berkeley~\cite{Bickel1975}. Analysis of the aggregate admissions data for the school revealed a statistically significant bias against women: a smaller fraction of female applicants were admitted. However, when admissions data was disaggregated by department, women were shown to have parity, and even a slight advantage in some departments, over men. The paradox arose because departments that female applicants prefer have lower admissions rates for both genders.

Computational social scientists are often interested in measuring relationships between some outcome and a covariate. Here too Simpson's paradox can distort results~\cite{Blyth1972,Lerman2018jcss}. For example, a study~\cite{Vaupel85heterogeneity} of recidivism showed that the rate at which released convicts return to prison declines over time. From this, one may conclude that age has a pacifying effect, and older convicts are less likely to re-offend. In reality, however, this is not the case. Instead, the population of ex-convicts can be considered to be composed of two subgroups with nearly constant, but different recidivism rate. The first subgroup is composed of convicts that have been reformed and will never commit a crime once they are released from prison. The second  subgroup is composed of incorrigibles, who are highly likely to re-offend. Over time, members of this subgroup commit offenses and return to prison, leaving fewer of them in the population. Survivor bias changes the composition of the population, giving an illusion of an overall decline in recidivism. As Vaupel and Yashin warn, ``unsuspecting researchers who are not wary of heterogeneity's ruses may fallaciously assume that observed patterns for the population as a whole also hold on the sub-population or individual level.'' Researchers have not routinely tested for the presence of Simpson's paradox in their data, although such as test was recently proposed~\cite{Lerman2018jcss}.

The idea for using Simpson's paradox to uncover interesting patterns in data was explored by researchers in the past~\cite{fabris2000discovering}. Our work extends previous research in new directions. We not only provide a principled way to disaggregate data, but also introduce a novel measure that allows us to quantify how interesting or surprising an instance of Simpson's paradox is. As we demonstrate in the results, this provides a novel tool for studying behavioral data.

% ______________________________________________________________________
%                                                 METHODS
% ______________________________________________________________________
\section{Methods} \label{sec:methods}
We describe \emph{Simpson's Disaggregation}, a method for automatically uncovering interesting patterns in data.
The method takes as input a set of observations of an outcome, $Y$, and a set of covariates, or features, $X = \{ X_1, X_2, ..., X_m \}$ associated with it.
The approach has three steps. First, it disaggregates data into more homogenous subgroups based on some covariate $X_c$. Next, it uses a linear model to capture trends with respect to some other covariate $X_j$, both within the subgroups  and within the aggregate data. Finally, it quantifies how well the models describe the disaggregated data compared to aggregate data to identify the important disaggregations. We describe these steps in detail below.\footnote{The code implementing the method is available on https://github.com/ninoch/Trend-Simpsons-Paradox/.}

\subsection{Step 1: Disaggregating Data}
We disaggregate the data by partitioning it on the conditioning variable $X_c$  into non-overlapping bins, such that data points within each bin are more similar to each other than to data in other bins. These bins correspond to the more homogeneous subgroups within the population generating the data. Simply partitioning the data into fixed-size bins ~\cite{Alipourfard2018}, or percentiles, can be problematic when $X_c$ has a heavy-tailed distribution, since the bins covering the tail will have few data points in them. In such cases, logarithmic binning is a better choice. However, the decision then has to be made about the size and scale of each bin. This decision must balance two considerations: first, each bin has to be homogeneous, i.e., it must contain data points that are more similar to each other in relation to the outcome variable than to variables in other bins, and secondly, it needs to have a sufficient number of data points. Basically, too small a bin may not contain enough samples for a statistically significant measurement, while the samples in too large a bin may be too heterogeneous for a robust trend.

The binning method described below partitions the values of $X_c$, such that $Y$ exhibits little variation within each bin but significant variation between bins.

\subsubsection{Quantifying the Partition}
Total sum of squares (SST) is the key concept used to describe the variation in observations $\{ y_i \}^{N}_{i = 1}$ of a random variable $Y$. It is defined as $SST = \sum_{i=1}^N (y_i - \bar{y})^2$, where $\bar{y} = \frac{1}{N} \sum_{i=1}^N y_i$ is the mean of all observations. The sample variance, $\sigma^2$, is equal to $SST / (N - 1)$, thus the SST is related to variation in $Y$.
For any arbitrary partition $P_{X_c}$ of the variable $X_c$, we can quantify how much variation of the outcome variable $Y$ can be explained by $P_{X_c}$ by decomposing the total sum of squares as:
\begin{equation}
\sum_{i=1}^N (y_i - \bar{y})^2 = \sum_{b \in P_{X_c}} N_b(\bar{y}_b - \bar{y})^2 + \sum_{b \in P_{X_c}} \sum_{i = 1}^{N_b} (y_{b, i} - \bar{y}_b)^2,
\label{eq:sst}
\end{equation}
where $N_b$ is the number of data points in bin $b$, $y_{b, i}$ is the $i$-th data point in bin $b$, and $\bar{y_b}$ is the average of values in that bin. The first term on the right hand side of Eq.~(\ref{eq:sst}) is the sum of squares between groups, a weighted average of squared differences between global ($\bar{y}$) and local ($\bar{y_b}$) average. This sum measures how much $Y$ varies between different bins of the partition. The second term is the sum of squares within groups, which measure how much variation in $Y$ remains within each bin $b$. Then, the proportion of the explained sum of squares to the total sum of squares, or coefficient of determination, is:
\begin{equation}
R^2 = \frac{\sum_{b \in P_{X_c}} N_b(\bar{y}_b - \bar{y})^2}{SST}
\label{eq:r_2}
\end{equation}
The $R^2$ measure takes values between zero and one, with large values of $R^2$ indicating a larger proportion of the variation of $Y$ explained by $X_c$, for this specific binning $P_{X_c}$.

\subsubsection{Finding the Best Partition}
Now, we will describe the systematic way for learning partition $P_{X_c}$ for the feature $X_c$, which explains the largest possible variation of the outcome $Y$. Given the data, the domain of the feature $X_c$ can be split at some value $s$ into two bins: $X_c \leq s$ and $X_c > s$. From Eq.~(\ref{eq:r_2}), the proportion of variation in $Y$ explained by such a split is:
\begin{equation}
R^2(s; X_c) = \frac{N_{b_1} (\bar{y}_{b_1} - \bar{y})^2 + N_{b_2} (\bar{y}_{b_2} - \bar{y})^2}{SST},
\label{eq:r_2_bins}
\end{equation}
where $N_{b_1}$ and $\bar{y}_{b_1}$ are the number of data points and average value of $Y$ in the bin $X_c \leq s$, and $N_{b_2}$ and $\bar{y}_{b_2}$ are the number of data points and average value of $Y$ in the bin $X_c > s$. The $s$ can take any value in the domain of $X_c$, and afterward the $R^2(s; X_c)$ can be computed for that $s$. Thus, among all possible values for $s \in [min(X_c), max(X_c)]$, we can choose $s_1$ as the optimal split for $X_c$ which maximizes $R^2(s; X_c)$. For the next iteration, we can choose the next best split $s_2$ to optimize improvement in $R^2$. In general, assume we have bins $\{b_u\}_{u=1}^k$ after $k - 1$ iterations, and for next iteration we have found best split $s_{k + 1}$ which divides the bin $b_i$ into two bins, $b_{i_1}$ and $b_{i_2}$ where $b_{i_1}$ associated with data points in bin $b_i$ where $X_c \leq s_{k + 1}$, and $b_{i_2}$ associated with data points in bin $b_i$ where $X_c > s_{k + 1}$. Thus, after splitting we will have bins $b_{i_1}$ and $b_{i_2}$ instead of bin $b_i$. In this case, the $R^2$ improvement is the following:
\begin{multline}
\begin{split}
\Delta R^2 (s | P_{X_c}; X_c)= \frac{1}{SST} \left( N_{b_{i_1}} (\bar{y}_{b_{i_1}})^2 + N_{b_{i_2}} (\bar{y}_{b_{i_2}})^2 \right. \\
\left. - N_{b_i} (\bar{y}_{b_i})^2 \right)
\end{split}
\end{multline}

In this manner, the method recursively splits the domain of $X_c$ to create a partition of the feature.
However, this procedure will continue indefinitely until $X_c$ has been partitioned into bins consisting of single points, overfitting the data.
To prevent this, we constrain the algorithm so that the maximum number of bins is 20, while the minimum number of data points per bin is 100.

\subsection{Step 2: Modeling Disaggregated Data}
Next, the method measures the association between the outcome variable $Y$ and the independent variable $X_j$ in the aggregate data and compares it to the associations in the disaggregated data.

At an aggregate level, we model the relationship between $Y$ and $X_j$ as a linear model of the form
\begin{equation}
	\mathbb{E}[Y|X_j = x_j] = f(\alpha + \beta X_j),
	\label{eq:f_p}
\end{equation}
where $f(\alpha + \beta X_j)$ is a monotonically increasing function of its argument $(\alpha + \beta X_j)$. The parameter $\alpha$ in Eq.~\eqref{eq:f_p} is related to the intercept of the regression function, while the coefficient $\beta $ quantifies the effect of $X_j$ on $Y$. For the disaggregated data, we fit linear models of the same form
but with different values of the parameters $\alpha$ and $\beta $ depending on values of the conditioning variable $X_c$:	
\begin{equation}
	\mathbb{E}[Y|X_j = x_j, X_c = x_c] = f(\alpha(x_c) + \beta (x_c)X_j).
	\label{eq:f_pc}
\end{equation}

To check whether disaggregating data on $X_c$ results in a Simpson's paradox, we look for trend reversal by comparing the sign of $\beta$ from the fit to the aggregate data (Eq.~\ref{eq:f_p}) to  the signs of $\beta$'s from fits to the disaggregated data (Eq.~\ref{eq:f_pc}) (given that $\beta$s are significantly different from zero).
if more than half of the subgroups have different sign with aggregated trend, then Simpson's paradox exists.

Trend reversal is an interesting phenomenon, because it occurs when subgroups exhibit behaviors that are different from the trends within the population as a whole. However, behavioral differences can exist even without trend reversal. Consider, for example, trend lines with zero slope that are stacked, because outcomes within each subgroup are systematically different from each other. In this case, disaggregating data is still desirable, even if the trend reversal rules for Simpson's paradox do not indicate it. Rather than simply look for trend reversal, we measure the significance of the disaggregations of data.

\subsection{Step 3: Significance of Disaggregations}
We conjecture that surprising subgroups are those whose behavior deviates substantially from that of the population as a whole.
Existence of such subgroups suggests that important behavioral differences exist that require deeper analysis. To identify such subgroups we must first quantify how well a model, in simplest case a linear model, describes the data.

In this paper, we examine the case where the outcome variable $Y$ is binary. In this case, $\mathbb{E}[Y|X_j = x_j]$ is the probability of $y_i = 1$ given $X_j = x_j$. Therefore, we use the logistic regression as our linear model, and Equation~(\ref{eq:f_p}) becomes:
\begin{equation}
\mathbb{E}[Y|X_j = x_j] = f(\alpha + \beta X_j) = \frac{1}{1 + e^{-(\alpha + \beta X_j)}}
\label{eq:7}
\end{equation}

Logistic regression uses Maximum Likelihood Estimation to find the best fit to data. Likelihood of the model $\mathcal{M}$ given the data is $\mathcal{L}(\mathcal{M} | x) = P(X = x | \mathcal{M})$. For a binary outcome variable, it becomes:
\begin{equation}
	\mathcal{L}(\mathcal{M} | x) = \prod_{i = 1}^{N} y_i \times (P_{\mathcal{M}}(x_i)) + (1 - y_i) \times (1 - P_{\mathcal{M}}(x_i))
	\label{eq:llg}
\end{equation}
and then the log likelihood is give by:
\begin{equation}
	log \mathcal{L}(\mathcal{M} | x) = \sum_{i = 1}^{N} y_i \times log(P_{\mathcal{M}}(x_i)) + (1 - y_i) \times log(1 - P_{\mathcal{M}}(x_i))
	\label{eq:lllg}
\end{equation}
For assessing the goodness of fit, we can use deviance~\cite{deviance}. It compares Log-likelihood of two models. In the case of the Logistic regression of Eq.~\ref{eq:7}, this corresponds to comparing the full model with a null model consisting only of an intercept.
Letting $\mathcal{M}_0$ be the reduced model and $\mathcal{M}_1$ the full model, the deviance of these two models is:
\begin{equation}
	D(\mathcal{M}_1, \mathcal{M}_0) = 2 \times [ log \mathcal{L}(\mathcal{M}_1 | x) - log \mathcal{L}(\mathcal{M}_0 | x) ]
	\label{eq:deviance}
\end{equation}
In the case where $\mathcal{M}_0$ is a nested model of $\mathcal{M}_1$, where nested means full model can be reduced to null model by imposing constraints on the parameters, under the null hypothesis that $\mathcal{M}_0$ and $\mathcal{M}_1$ provide a similar quality statistical explanations of the outcome,
for sufficiently large sample size, the deviance comes from a $\chi^2(p)$ distribution~\cite{deviance}, where $p$ is the number of degrees of freedom, which equals the number of extra parameters of $\mathcal{M}_1$ in comparison to $\mathcal{M}_0$. If the hypothesis is rejected, it means that $\mathcal{M}_1$ provides a significantly better description of the outcome variable than $\mathcal{M}_0$.

\subsubsection{Significance of Aggregate Data Model}
For assessing the significance of a model of aggregate data, we use deviance to compare the aggregate data model, given by  Eq.~(\ref{eq:f_p}), to a model where all $\hat{y_i}$ are equal to the global mean $\bar{y}=\frac{\sum_{i = 1}^N y_i}{N}$. In this case, Eq.~(\ref{eq:deviance}) becomes:
\begin{equation}
	D(\mathcal{M}_1, \mathcal{M}_0) = 2 \times \sum_{i = 1}^{N} y_i \times log(\frac{\hat{y_i}}{\bar{y}}) + (1 - y_i) \times log(\frac{1 - \hat{y_i}}{1 - \bar{y}})
\end{equation}

Where, $y_i$ is the $i$-th outcome, and $\hat{y_i}=f(\alpha + \beta x_i)$. Clearly, these two models are nested; therefore, $D(\mathcal{M}_1, \mathcal{M}_0)$ has a $\chi^2(1)$ distribution~\cite{deviance}. We can apply statistical hypothesis test to see whether
the found trend for aggregated data is significant or not.

\subsubsection{Significance of Disaggregated Data Model}
For assessing the significance of a disaggregation of data, we can use deviance to compare the model of Eq.~(\ref{eq:f_pc}) with a model where $\hat{y_i}$ is equal to the average outcome for data points in the bin of $x_i$. In this case, Eq.~(\ref{eq:deviance}) becomes:
\begin{equation}
	2 \times \sum_{b \in P_{X_c}} \sum_{i = 1}^{N_b} y_{b, i} \times log(\frac{\hat{y}_{b, i}}{\bar{y}_b}) + (1 - y_{b, i}) \times log(\frac{1 - \hat{y}_{b, i}}{1 - \bar{y}_b}),
\end{equation}
where, $y_{b, i}$ is the $i$-th data point in bin $b$, $\bar{y}_b=\frac{\sum_{i = 1}^{N_b} y_{b, i}}{N_b}$ is the mean outcome within bin $b$, and
$\hat{y}_{b, i}=f(\alpha(x_{b, i}^c) + \beta (x_{b, i}^c) \times x_{b, i}^j)$. By imposing $\beta(x_c) = 0, \forall x_c \in X_c$,
we conclude that these two models are nested. Thus again, we can use statistical test $\chi^2(|P_{X_c}|)$ to see whether
the disaggregated trends are significant or not.

\subsubsection{Comparing Disaggregations}
Comparing disaggregations of data based on how well the linear models describe trends within subgroups can help us identify interesting behavioral patterns in data. A disaggregation on variables $(X_{j_1}, X_{c_1})$ is more interesting than $(X_{j_2}, X_{c_2})$ if it has more explanatory power than the second pair. McFadden~\cite{mcfadden1}, introduced a measure, called \textit{McFadden $R^2$} or pseudo-$R^2$, to capture the ratio of likelihood improvement: % Page 121 of Analysis of qualitative choice behavior book
\begin{equation}
R^2_{McFadden} = 1 - \frac{log \mathcal{L}_{full}}{log \mathcal{L}_{null}}
\label{eq:mcr2}
\end{equation}
If we assume that the full model is at least good as the null model (means $log \mathcal{L}_{full} > log \mathcal{L}_{null}$), then the value of $R^2_{McFadden}$ is between zero and one, with larger values showing more improvement in log-likelihood, and values $0.2$ to $0.4$ considered to represent excellent fits~\cite{mcfadden2}. Thus, we can fix the null model and compute the value of $R^2_{McFadden}$ for all disaggregations. For the null model, we choose simple global average for all $Y$. Thus, the right hand side of Eq.~(\ref{eq:mcr2}) becomes: % Page 307 of Chapter 13 of  Bahvioural Travel Modelling book
\begin{equation}
1 - \frac{\sum_{b \in P_{X_c}} \sum_{i = 1}^{N_b} y_{b, i} \times log(\hat{y_{b, i}}) + (1 - y_{b, i}) \times log(1 - \hat{y_{b, i}})}{ \sum_{i = 1}^{N} y_i \times log(\bar{y}) + (1 - y_i) \times log(1 - \bar{y}) }
\end{equation}

We use pseude-$R^2$ to rank disaggregations by their explanatory power.
In addition, we can also use it to identify the best conditioning variable $X_c$ for disaggregating the data that best explains the trends with respect to a covariate $X_j$.

% ______________________________________________________________________
%                                                 RESULTS
% ______________________________________________________________________
\section{Results}
\label{sec:results}
We illustrate proposed method by applying it to study human performance data from several online domains.

\subsection{Stack Exchange}
First, we study answerer performance on Stack Exchange (SE).  Launched in 2008 as a forum for asking computer programming questions, Stack Exchange has grown to encompass a variety of technical and non-technical topics. Any user can ask a question, which others may answer. Users can vote for answers they find helpful, but only the asker can {accept one of the answers as the best answer to the question}.
We used anonymized data representing all answers to questions posted on Stack Exchange from August 2008 until September 2014.\footnote{\url{https://archive.org/details/stackexchange}} Approximately half of the 9.6M questions had an
accepted answer, and we included in the study questions that received two or more answers.

To understand factors affecting user performance on SE, we study the relationship between the various features extracted from data and the outcome, here a binary attribute denoting whether the answer written by a user is \emph{accepted by the asker as best answer to his or her question}.
To this end, for each answer written by a user, we create a list of features describing the answer and the user. Features include the numbers of \emph{words}, \emph{hyperlinks}, and \emph{lines of code} the answer contains, and its \emph{Flesch readability} score~\cite{Readability}. Features describing answerers are their \emph{reputation}, \emph{tenure} on SE (in seconds and in terms of \emph{percentile} rank) and the total \emph{number of answers} written during their tenure. These features relate to user experience. We also use activity-related features, including \emph{time since previous answer} written by the user, \emph{session length}, giving the number of answers user writes during the session, and \emph{answer position} within that session. We define a session as a period of activity without a break of 100 minutes of longer. %~\cite{Singer2016plosone}.

\begin{table}
  \caption{Variables defining important disaggregations of Stack Exchange data, along with their pseudo-$R^2$ scores.}
  \label{tab:SE_pairs}
  \begin{center}
\scalebox{0.85}{
  \begin{tabular}{|c|c|c|}
    \hline
    \textbf{$R^2_{Mc}$}&\emph{Covariate} \textbf{$X_j$}&\emph{Conditioning on} \textbf{$X_c$}\\% &\textbf{$\mathcal{S}\%$}\\
    \hline
    \hline
    0.03 & Answer position & Number of answers \\% & 85 \\
    0.03 & Session length & Number of answers \\% & 85 \\
    0.02 & Number of answers & Reputation \\% & 82 \\
    0.02 & Answer position & Reputation \\% & 82 \\
    0.02 & Session length & Reputation \\% & 82 \\
    0.01 & Readability & Lines of codes \\% & 78 \\
    $<10^{-2}$  & Answer position & Session length \\% & 100 \\
    $<10^{-2}$   & Time since prev ans & Answer position \\% & 83 \\
    \hline
\end{tabular}
}
\end{center}
\end{table}
Of the 110 potential disaggregations of SE data arising from all possible pairs of covariates, our method identified 8 as significant. Table~\ref{tab:SE_pairs} ranks these disaggregations %feature pairs
along with their pseudo-$R^2$ scores.
Note that user experience, either in terms of the \emph{reputation} or the \emph{number of answers} written by the user over his or her tenure, comes up as an important conditioning variable in several disaggregations. Features related to user activity, such as \emph{answer position} within a session, \emph{session length}, and \emph{time since previous answer}, appear as important dimensions of performance.
This suggests that answerer behavior over the course of a session changes significantly, and these changes are different across different sub-populations. %\cite{Ferrara2017dynamics}

\begin{figure*}[h!]
\begin{tabular}{@{}l@{}l@{}}
   \includegraphics[width=0.5\linewidth]{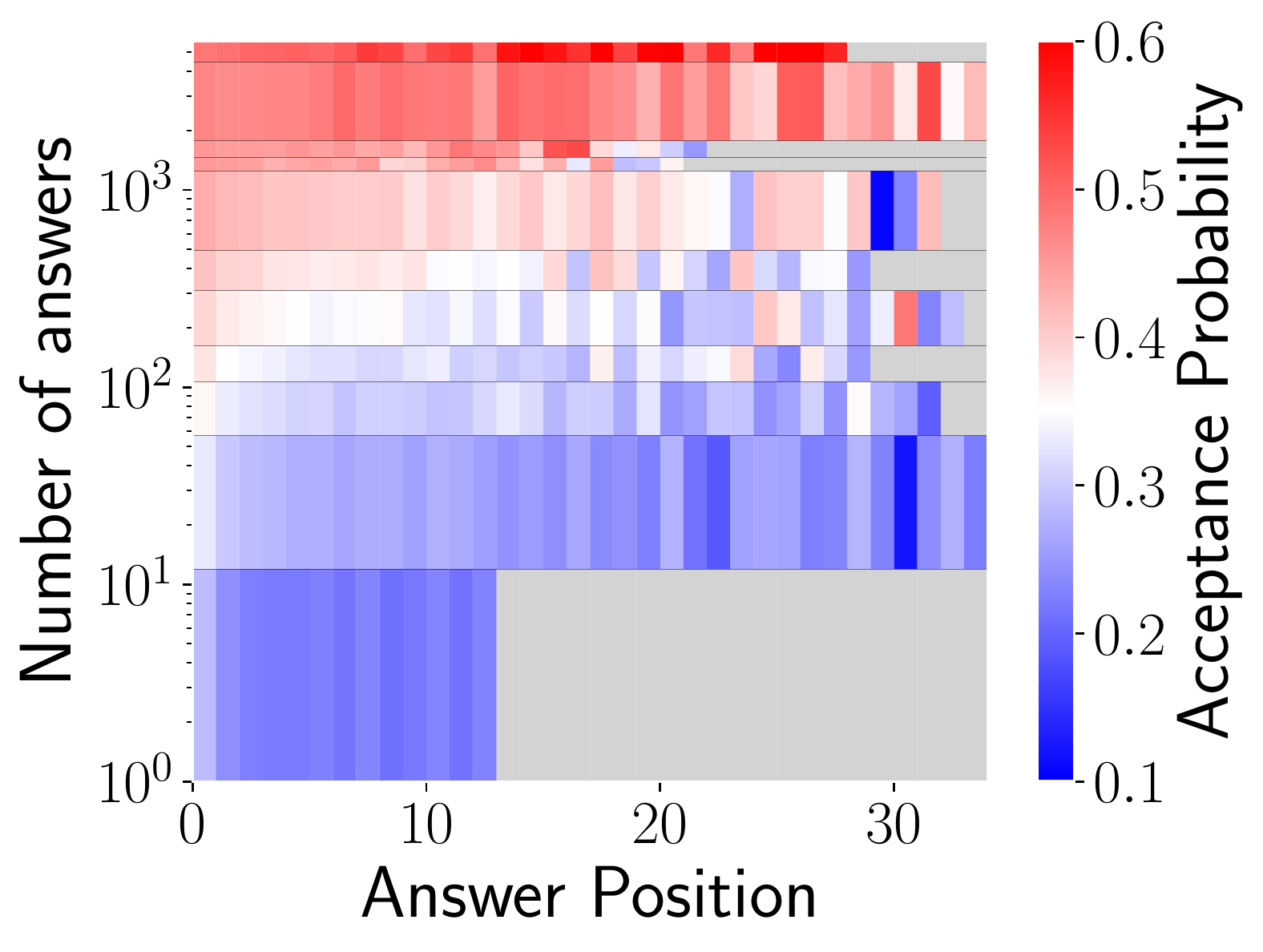} &
   \includegraphics[width=0.5\linewidth]{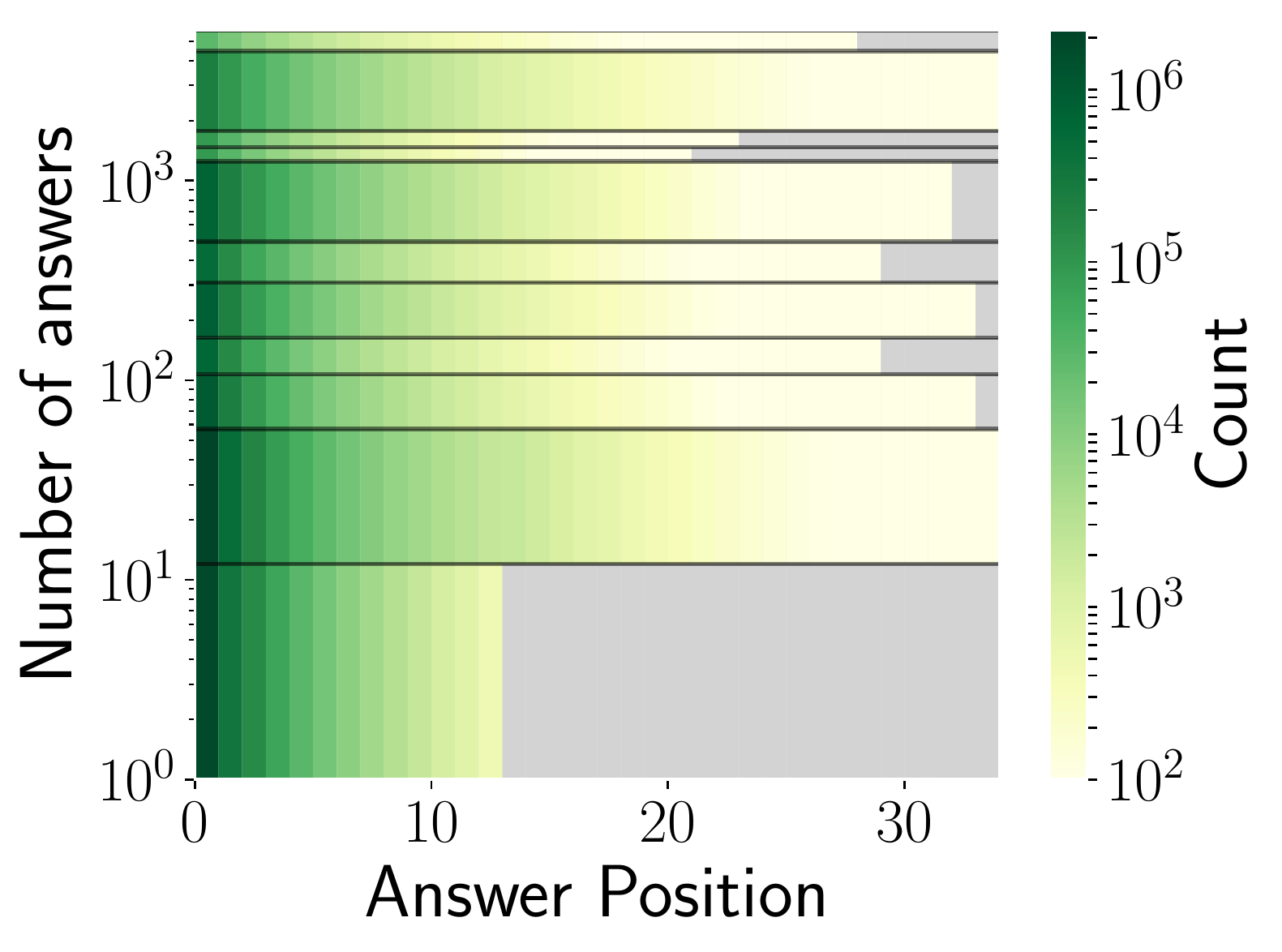} \\
  (a) Disaggregated data & (b) Number of samples \\
   \includegraphics[width=0.45\linewidth]{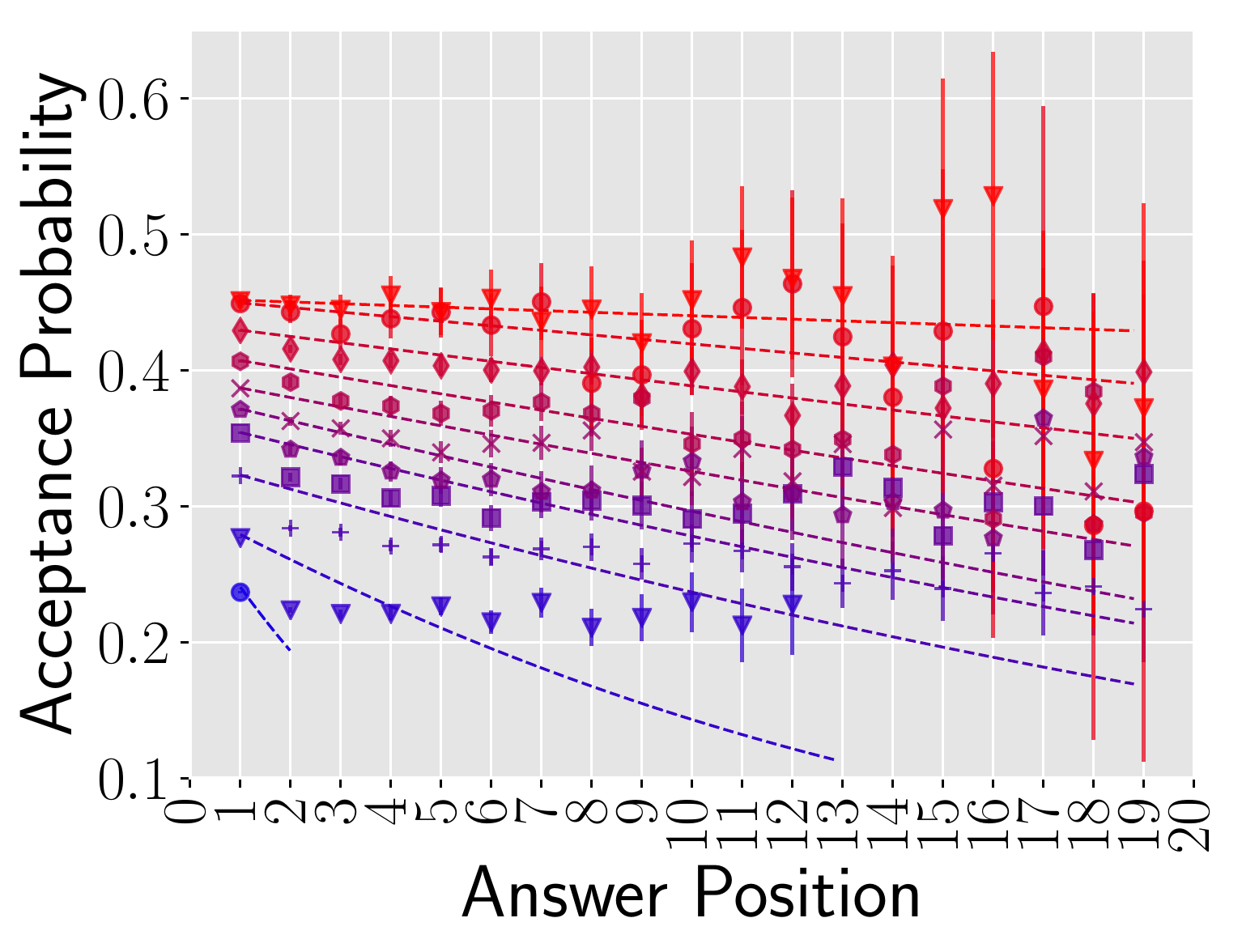} &
   \includegraphics[width=0.45\linewidth]{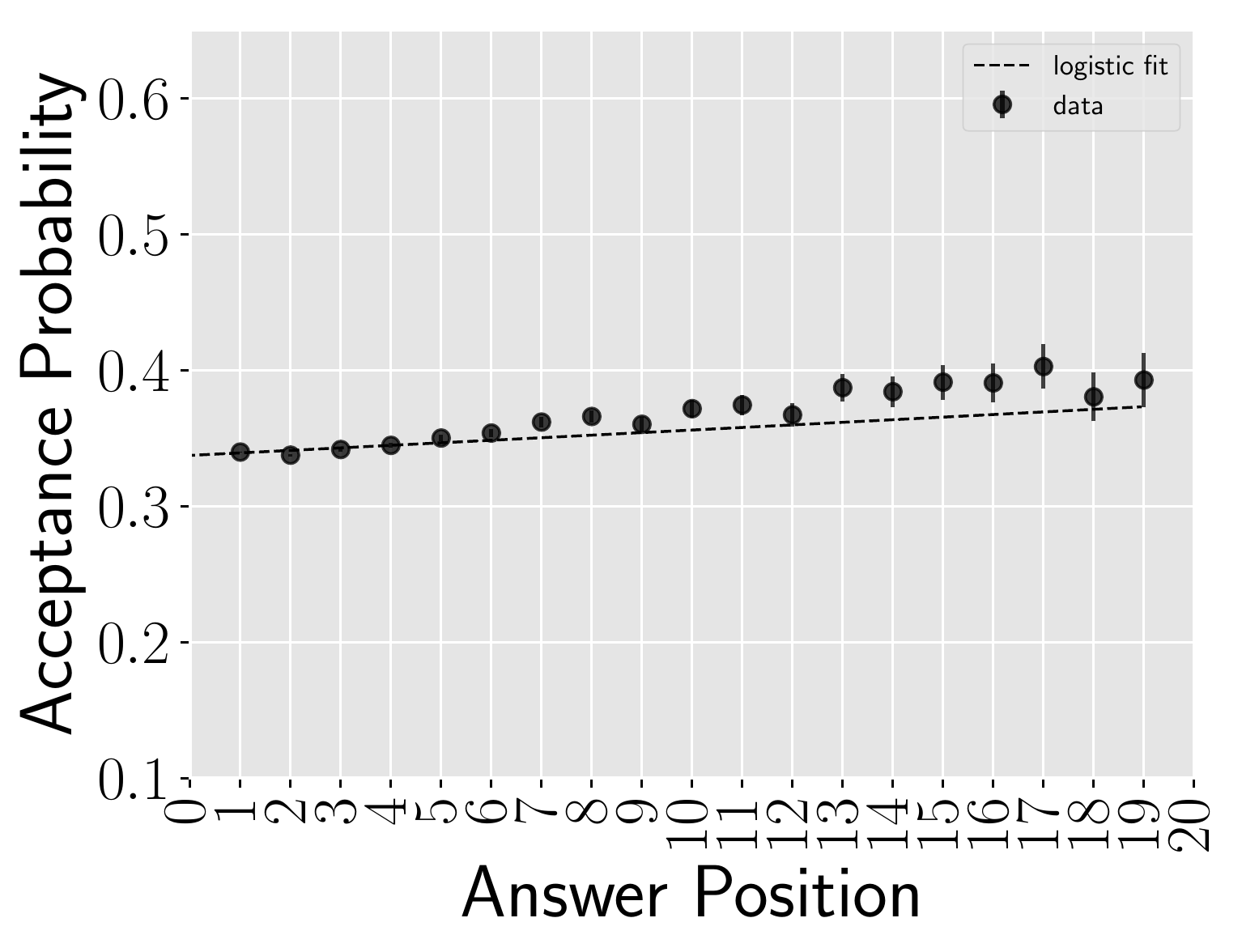} \\
   (c) Subgroup trends & (d) Aggregate trend
\end{tabular}
\caption{Disaggregation of Stack Exchange data. (a) The heat map shows the probability the answer is accepted as a function of its \emph{answer position} within a session, with the horizontal bands corresponding to the different subgroups, conditioned on total \emph{number of answers} the user has written. (b) Number of data samples within each bin of the heat map. Note that the outcome becomes noisy when there are few samples.
The trends in performance as a function of \emph{answer position} in (c) disaggregated data and (d) aggregate data. Error bars in (c) and (d) show 95\% confidence interval.
\label{fig:SE_paradox}}
\end{figure*}

Figure~\ref{fig:SE_paradox} visualizes the data, disaggregated on the number of answers. Each horizontal band in the heatmap in Fig.~\ref{fig:SE_paradox}(a) is a different bin of the conditioning variable \emph{number of answers}, and it corresponds to a distinct subgroup within the data. The first bin ranges in value from one to eleven answers, the second bin from 12 to over 50 answers, etc. Within each bin, the color shows the relationship between the outcome---the probability the answer is accepted---and \emph{answer's position} within a session. Dark blue corresponds to the lowest acceptance probability, and dark red to the highest. Within each bin, the color changes from lighter blue to darker blue (for the bottom-most bins), indicating a lower acceptance probability for answers written later in the session. For the top-most bins, the acceptance probability is overall higher, but also decreases, e.g., from pink to white to blue. Note that data is noisy, as manifested by color flipping, where there are few data points  (Fig.~\ref{fig:SE_paradox}(b)).

The trends corresponding to these empirical observations are captured in Fig.~\ref{fig:SE_paradox}(c). Note that the decreasing trends are in contrast to the trend in aggregate data (Fig.~\ref{fig:SE_paradox}(d)), which shows performance increasing with \emph{answer position} within the session. This suggests that user experience, as captured by the number of answers, is an important factor differentiating the behavior of users.

\begin{figure*}[h!]
\centering
\begin{tabular}{@{}l@{}l@{}}
   \includegraphics[width=0.5\linewidth]{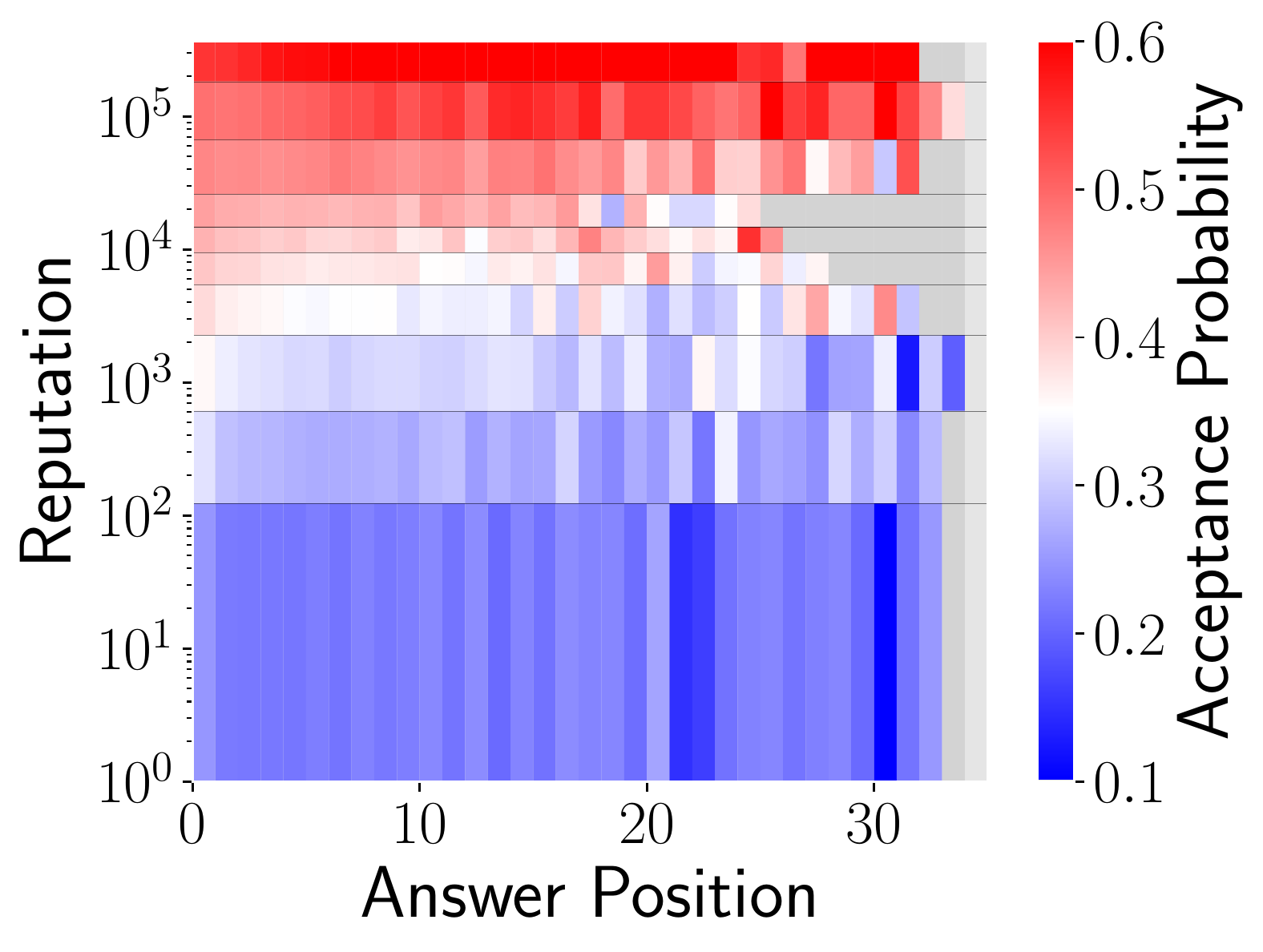} &
   \includegraphics[width=0.5\linewidth]{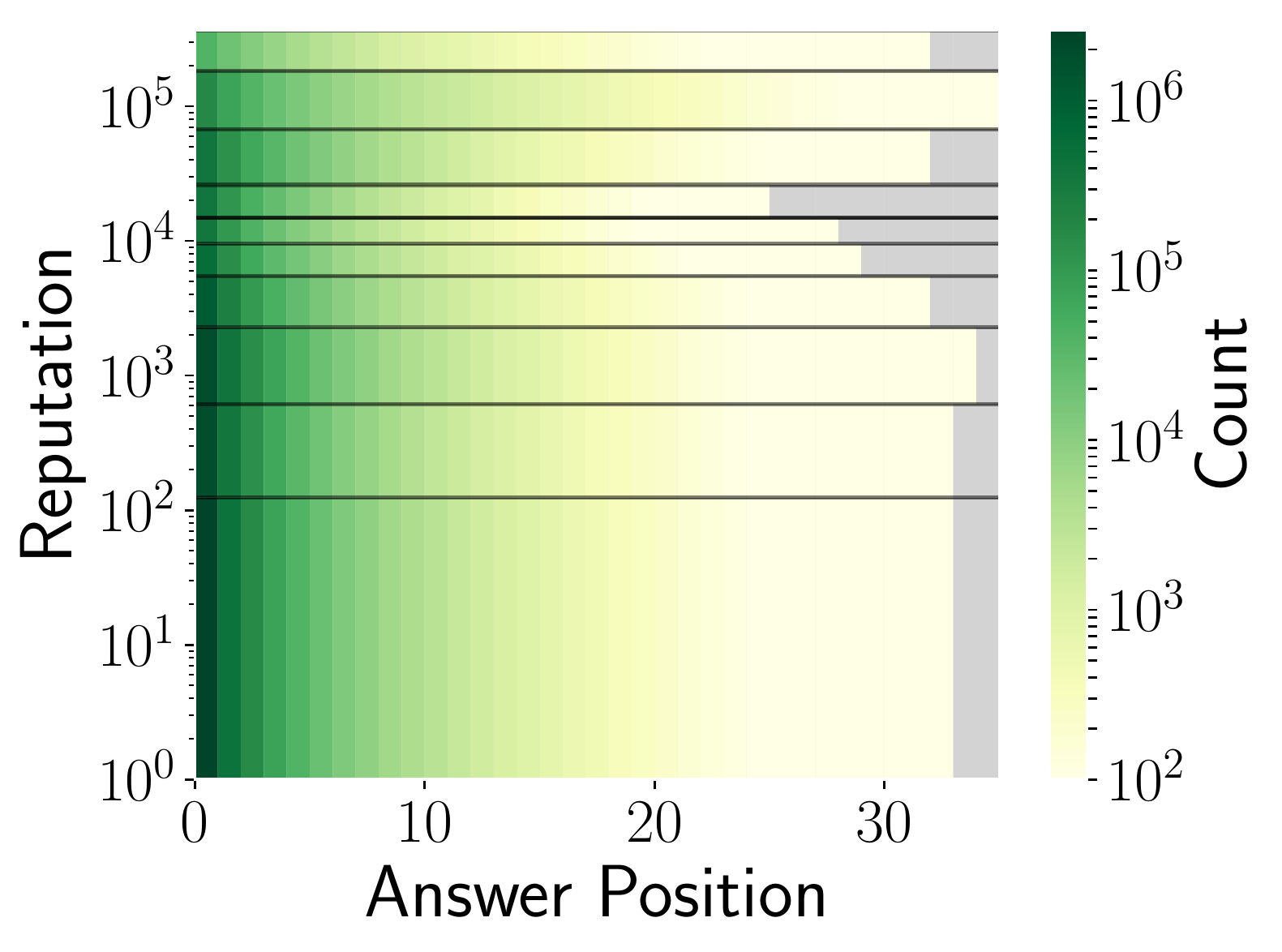} \\
  (a) Disaggregated data & (b) Number of samples \\
   \includegraphics[width=0.45\linewidth]{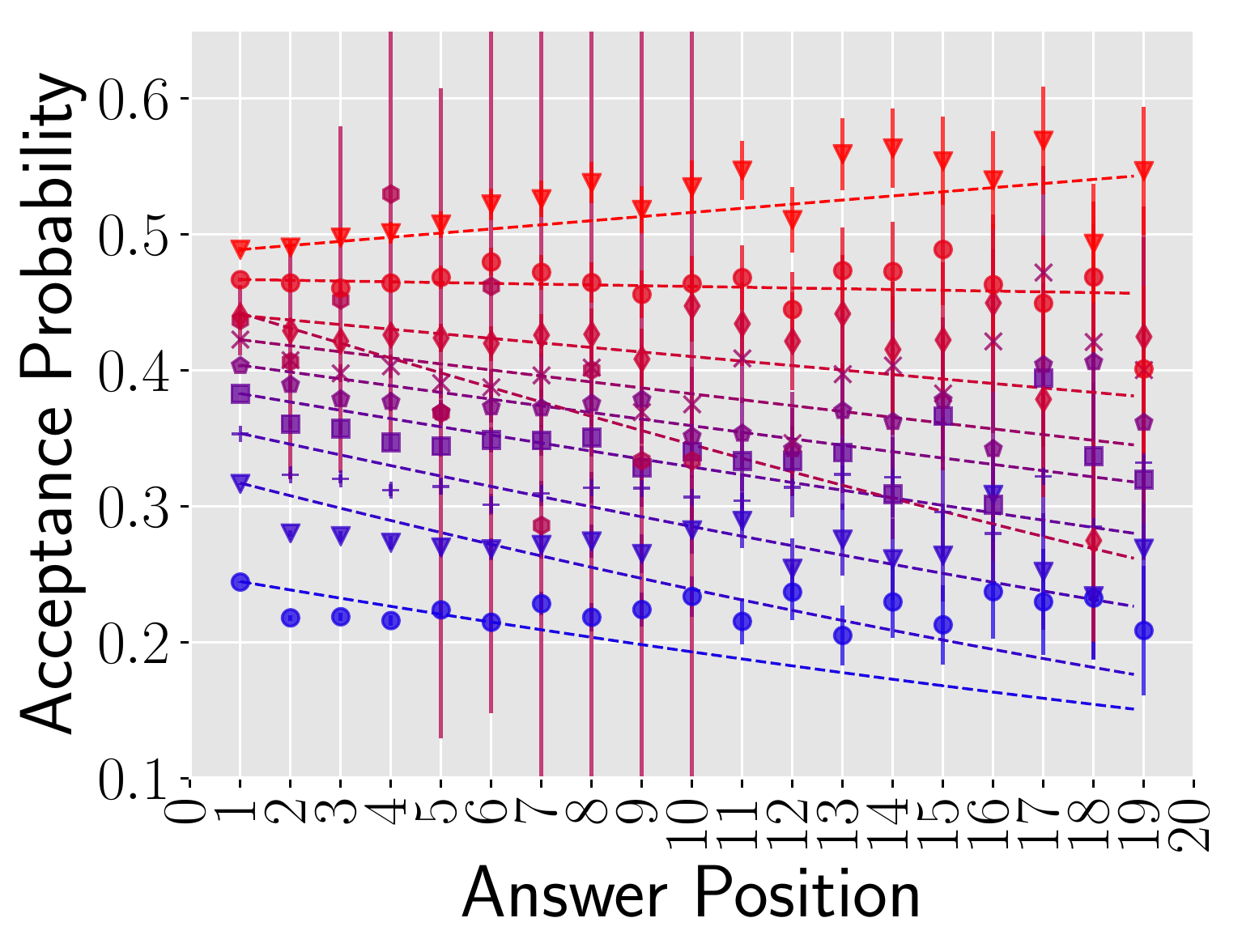} &
   \includegraphics[width=0.45\linewidth]{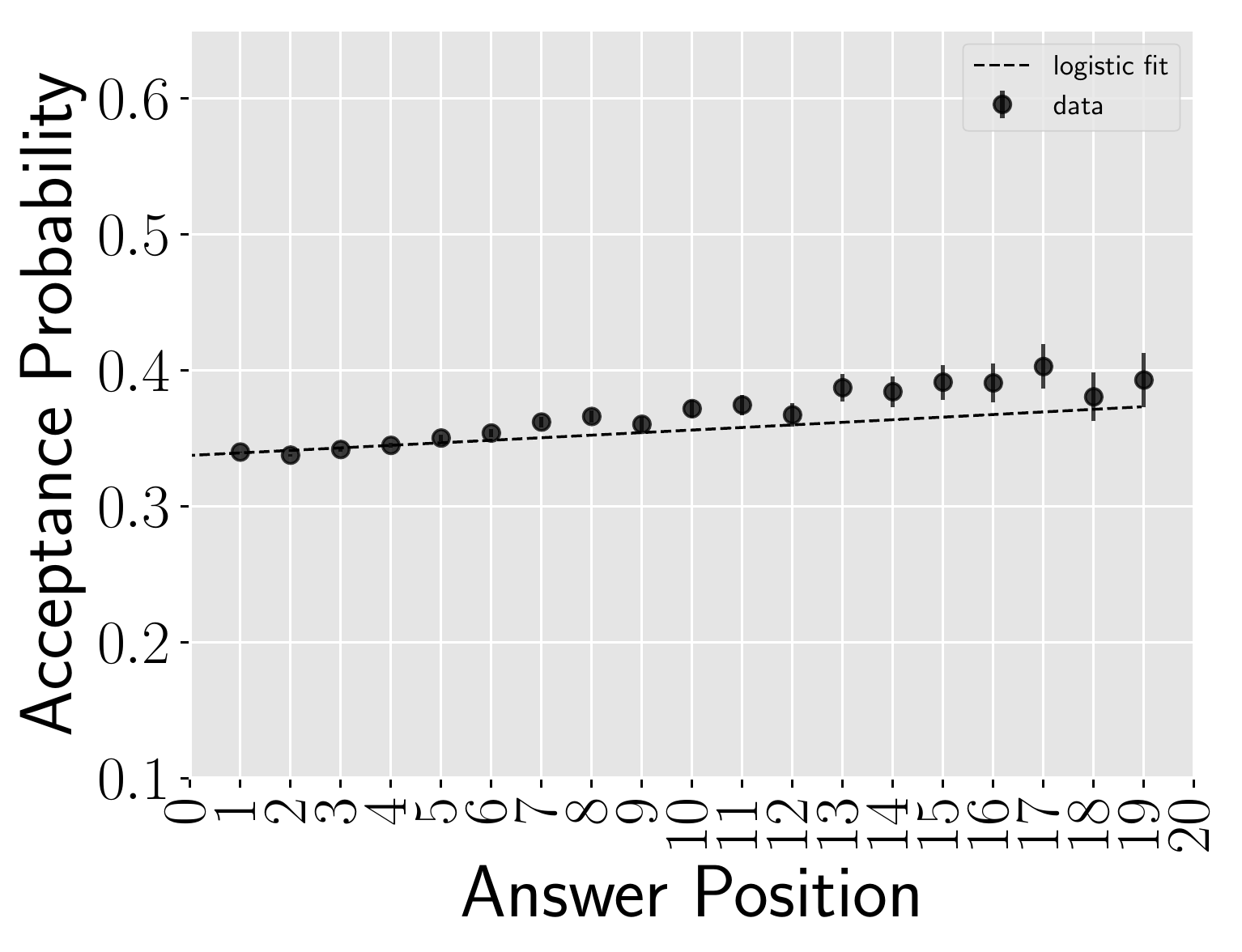} \\
   (c) Subgroup trends & (d) Aggregate trend
\end{tabular}
\caption{Disaggregation of Stack Exchange data similar to Fig.~\protect\ref{fig:SE_paradox}, but instead disaggreagted on user \emph{reputation}. (a) The heat map shows acceptance probability as a function of its \emph{answer position} within a session. (b) Number of data samples within each bin of the heat map. Note that the outcome becomes noisy when there are few samples.
The trends in (c) disaggregated data and (d) aggregate data.% Error bars in (c) and (d) show 95\% confidence interval.
\label{fig:SE_paradox2}}
\end{figure*}

Figure~\ref{fig:SE_paradox2} shows an alternate disaggregation of SE data for the covariate \emph{answer position}, here conditioned on user \emph{reputation}. This disaggregation is slightly worse, resulting in a somewhat lower value pseudo-$R^2$. While performance declines in the lower reputation subgroups as a function of answer position, the highest reputation users appear to write better answers in longer sessions.
The acceptance probability for high reputation users is more than 0.50, potentially indicating that askers pay attention to very high reputation users and are more likely to accept their answers.

\subsection{Khan Academy}

Khan Academy\footnote{https://www.khanacademy.org} (KA) is an educational platform offering online tools to help students learn a variety of subjects.  Student progress by watching short videos and complete exercises by solving problems. We studied an anonymized dataset, collected over two years, which contains information about attempts by KA adult students to solve problems. We partitioned student activity into sessions, also defined as a sequence of problems without a break of more than one hour between them. The vast majority of students completed only a single session.

As an outcome variable in this data, we take student performance on a problem, a binary variable equal to one when the student solved the problem correctly on the first try, and zero otherwise (either did not solve it correctly, or used hints).  To study factors affecting performance, we extracted the features of problems and users. These included the \emph{overall solving time} during user activity, \emph{total solve time} and the \emph{number of attempts} made to solve the problem, \emph{time since the previous problem} (tspp), the \emph{number of sessions} prior to the current one, \emph{all sessions} user contributed to, the \emph{session length} in terms of the number of problems solved, \emph{problem position} within the session (session index of the problem), the \emph{timestamp} of the attempt, including the \emph{month}, \emph{day of week}, \emph{type of weekday} ( whether it is weekend or not) and \emph{hour} the student attempted to solve the problem, the \emph{month the student joined} KA, his or her \emph{tenure}, the number of \emph{all attempts} on \emph{all problems} solved since joining, and how many of the problems were solved correctly on the \emph{first attempts}. As a proxy of skill or some background knowledge the student brings, we use how many problems were correctly solved during the student's \emph{five first attempts} to solve problems. For example, the least prepared students answered few of the five problems they attempted to solve, but best students would have solved all five correctly.

\begin{table}
  \caption{Variables defining important disaggregations of the Khan Academy data, along with their pseudo-$R^2$ scores.}
  \label{tab:ka_disagg}
\begin{center}
\scalebox{0.85}{
  \begin{tabular}{|c|c|c|}
    \hline
    \textbf{$R^2_{Mc}$}& \emph{Covariate} \textbf{$X_j$}&\emph{Conditioning on }\textbf{$X_c$}\\% &\textbf{$\mathcal{S}\%$}\\
    \hline
    \hline
0.06 & All  attempts& All first attempts \\
0.03 & All  attempts& All problems \\
0.01 & All  attempts& Tenure \\
0.01 & All  attempts& Total solve time \\
\hline
0.04 & Hour24 & All first attempts \\
\hline
0.04 &Session number & All first attempts \\
0.02 &Session number & All problems\\
0.01 &Session number & Tenure\\
0.01 &Session number & All  attempts\\
0.01 &Session number & All sessions\\
0.01 &Session number & Total solve time\\
0.0 &Session number & Join month\\
\hline
0.03 & Month & Five first attempts  \\
0.01 & Month & Session index \\
0.01 & Month & Total solve time \\
0.01 & Month & Timestamp \\
$<10^{-2}$ & Month & Week day \\
    \hline
0.01 & Problem position & Session length \\
\hline
\end{tabular}
}
\end{center}
\end{table}

Our method identified 32 significant disaggregations of KA data, out of 342 potential disaggregations. Some of these  are presented in Table~\ref{tab:ka_disagg}. The table lists conditioning variables for selected covariates, sorted by their pseudo-$R^2$ scores. For example, when examining how performance---probability to solve a problem correctly---changes over the course of a day ($X_j$ is \emph{hour24}), the relevant disaggregation conditions the data on \emph{all first attempts}, i.e., the number of all problems the user solved correctly on their first attempt. On the other hand, several disaggregations can explain the trends in performance as a function of \emph{month}. Conditioning on \emph{first five attempts} has the most explanatory power, followed by disaggregations conditioned on \emph{session index}, the \emph{total time} it took the user to solve all problems, the \emph{timestamp} and \emph{weekday} of the attempt.
Many of the conditioning variables used in the disaggregations represent different aspects of user experience on the site: the number of problems they tried to solve or correctly solved, their tenure on the site, and how much time they spent solving problems.

\begin{figure}[h!]
\centering
\begin{tabular}{@{}l@{}l@{}}
  \includegraphics[width=0.5\linewidth]{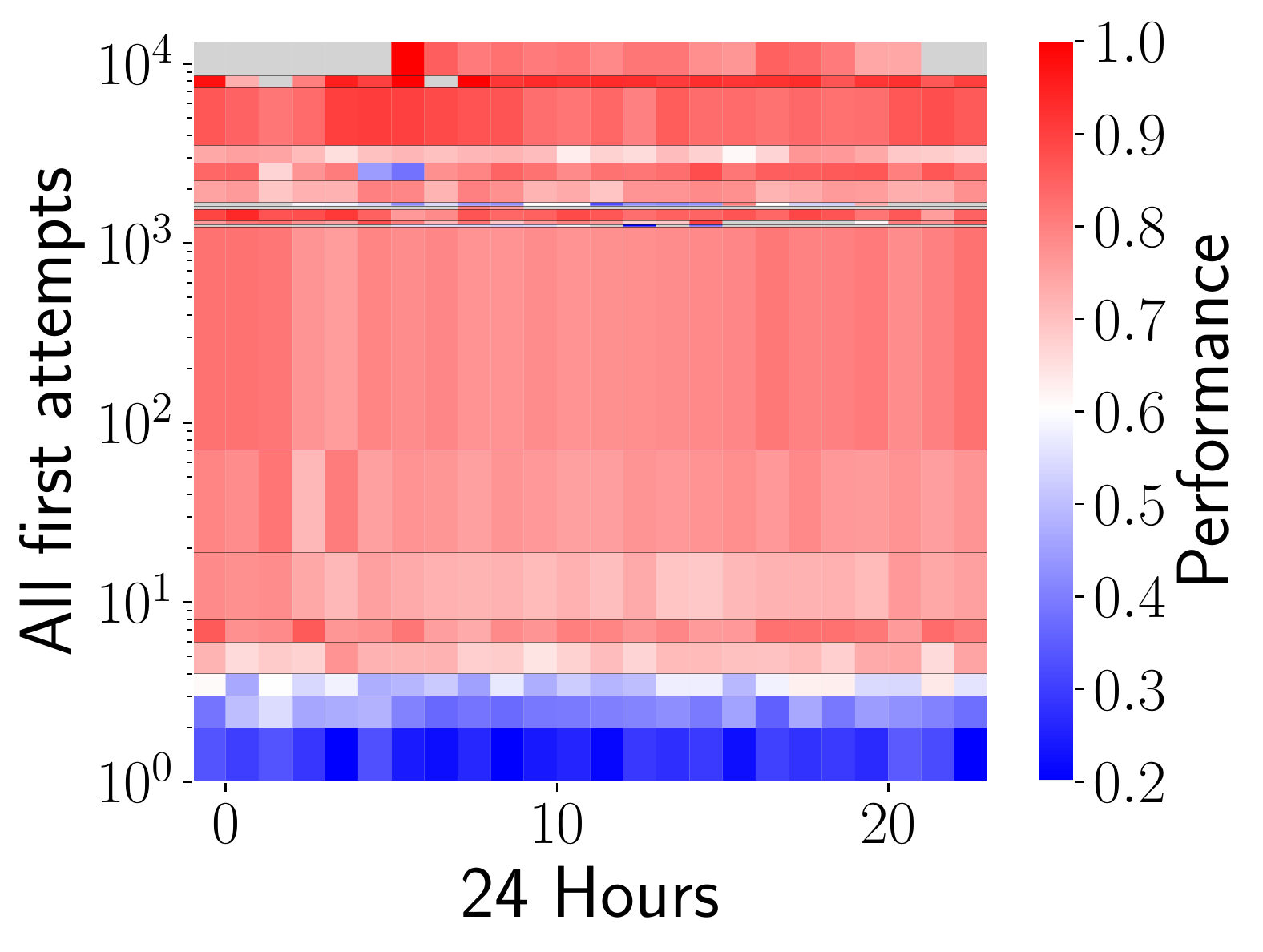} &
  \includegraphics[width=0.5\linewidth]{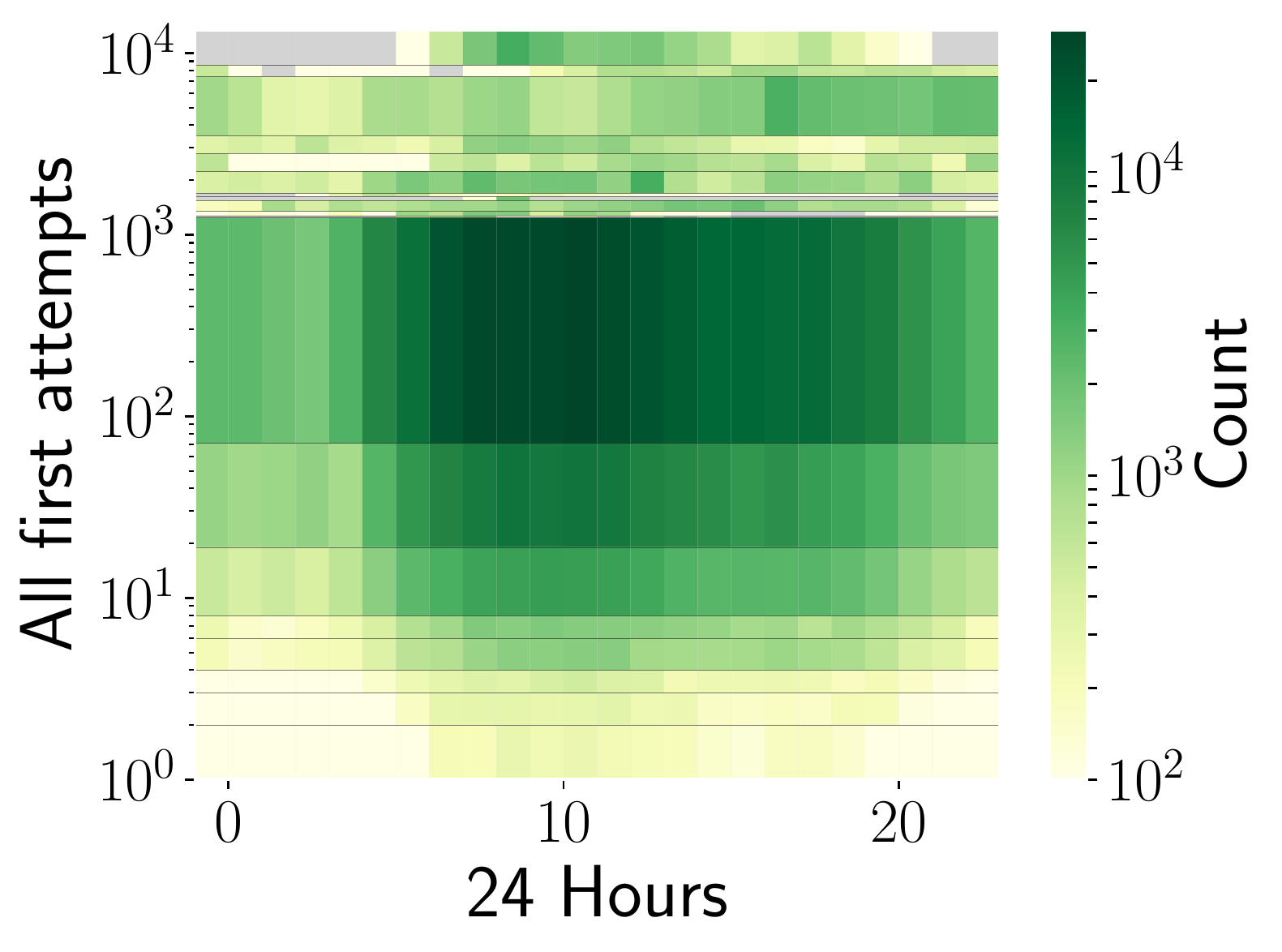} \\
  (a) Disaggregated data & (b) Number of samples \\
  \includegraphics[width=0.45\linewidth]{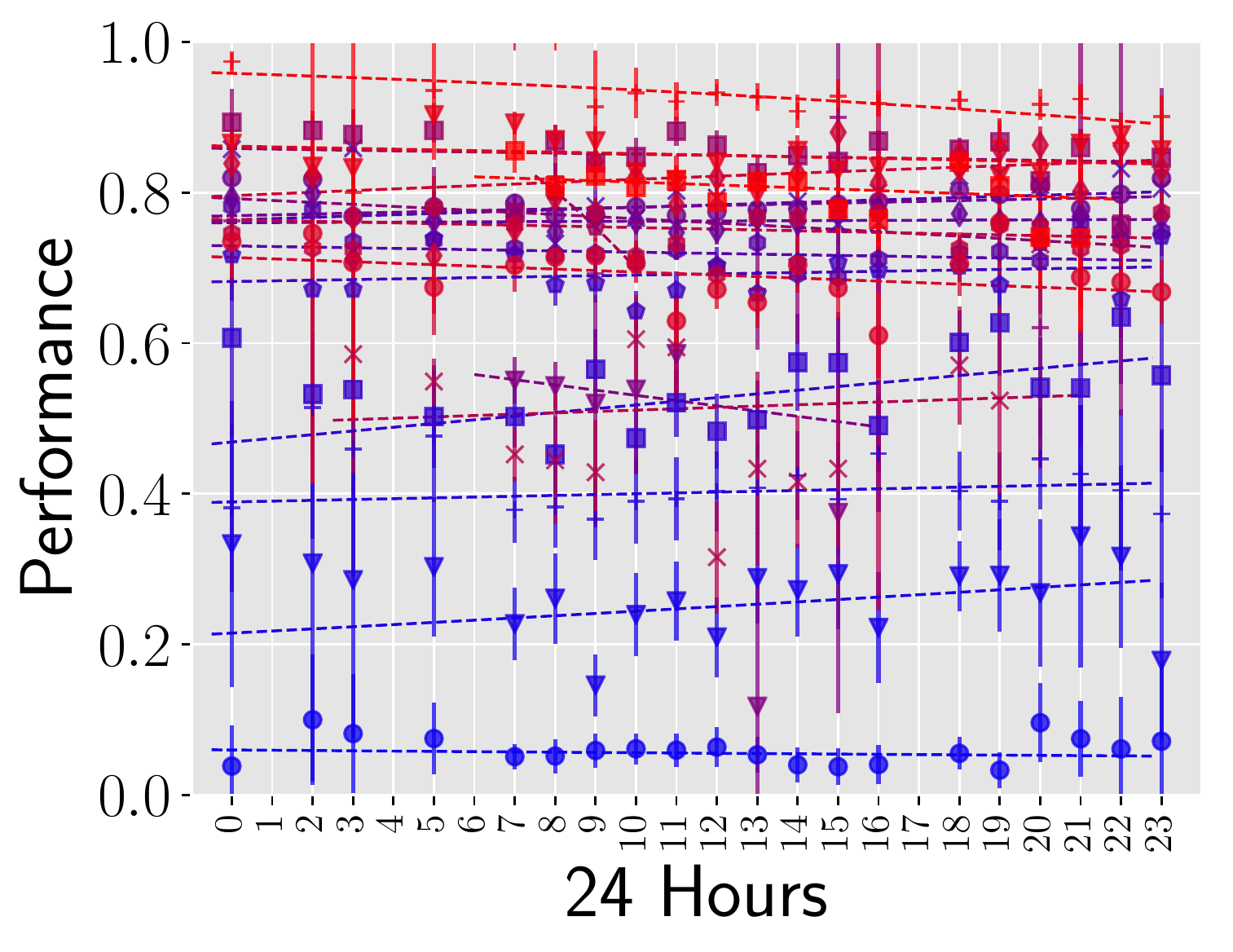}  &
  \includegraphics[width=0.45\linewidth]{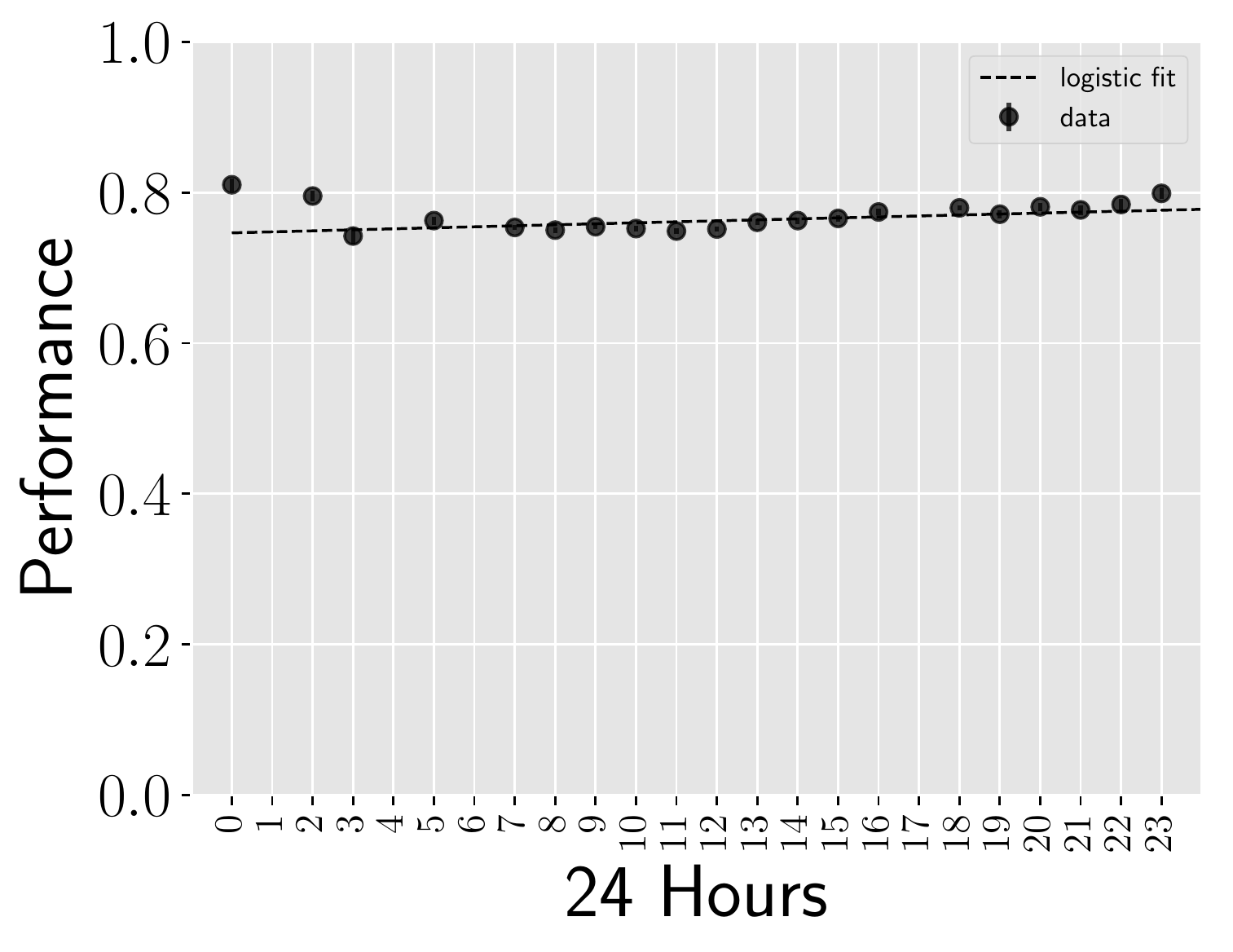} \\
  (c) Subgroup trends & (d) Aggregate trend
\end{tabular}
\caption{A disaggregation of the Khan Academy data showing performance as a function of \emph{hour} of day, conditioned on \emph{all first attempts}. (a) The heat map shows average performance within a subgroup as a function of the \emph{hour} of day.  (b) Number of data samples within each subgroup.
The trends in (c) the disaggregated data and in (d) aggregated data. %Error bars in (c) and (d) show the 95\% confidence interval.
\label{fig:ka2_paradox}}
\end{figure}

Figure~\ref{fig:ka2_paradox} takes a closer look at the disaggregation corresponding to covariate \emph{hour24}. In the aggregate data (Fig.~\ref{fig:ka2_paradox}(d)), there is a small but significant upward trend in performance over the course of a day. It looks like performance is higher at night than during the day. However, when data is disaggregated by \emph{all first attempts}, only a couple of subgroups have the up-trend: the rest stay flat or even decline in performance. All first attempts, which represents how many of all problems users solved correctly on their first try, captures both user's motivation to use KA (the more motivated, the more problems they attempt), and skill (the more skilled, the more problems they will solve on their first attempt). The high-achieving users actually perform better in the morning, in contrast to aggregate trends.

\begin{figure}[h!]
\centering
\begin{tabular}{@{}l@{}l@{}}
  \includegraphics[width=0.5\linewidth]{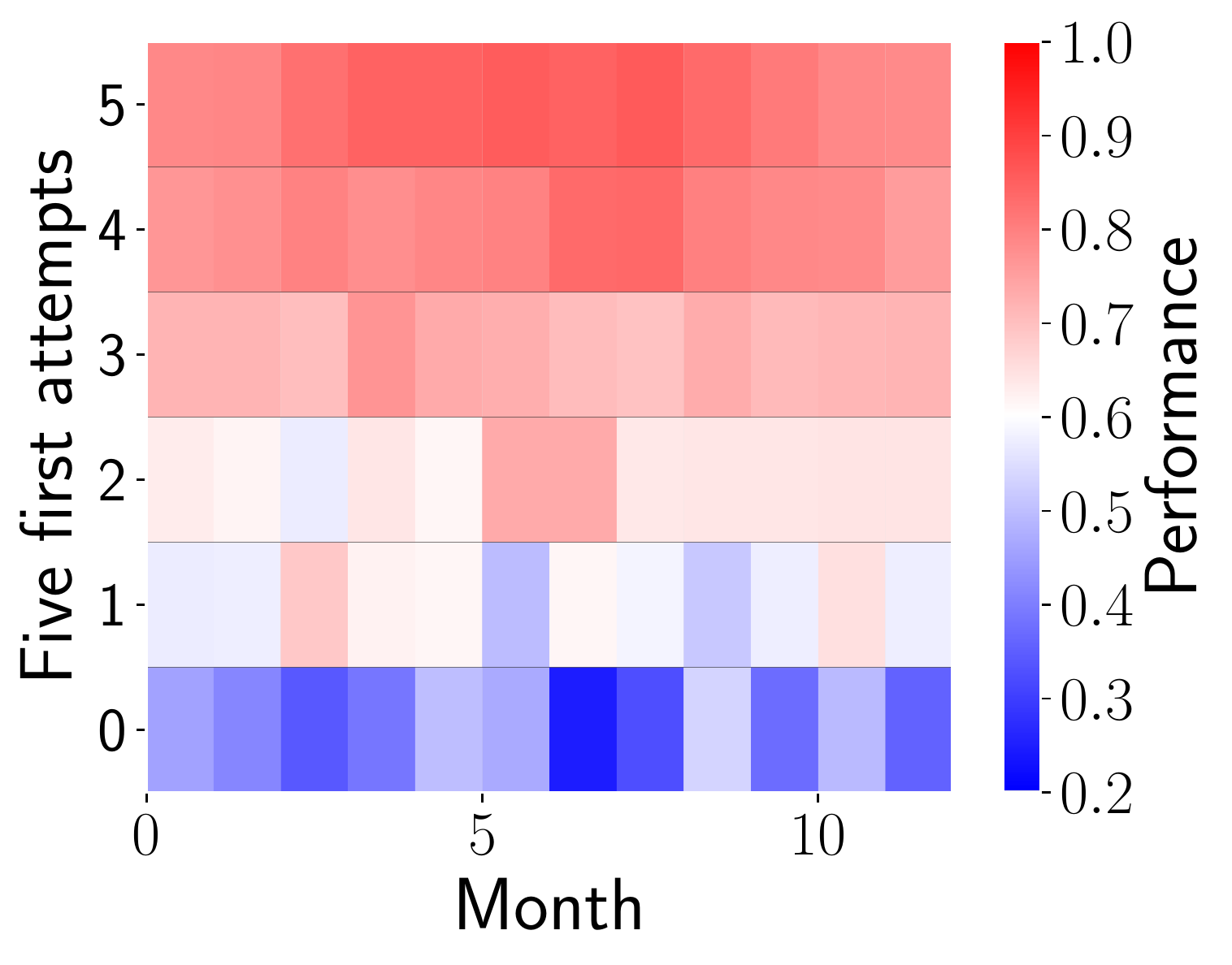} &
  \includegraphics[width=0.5\linewidth]{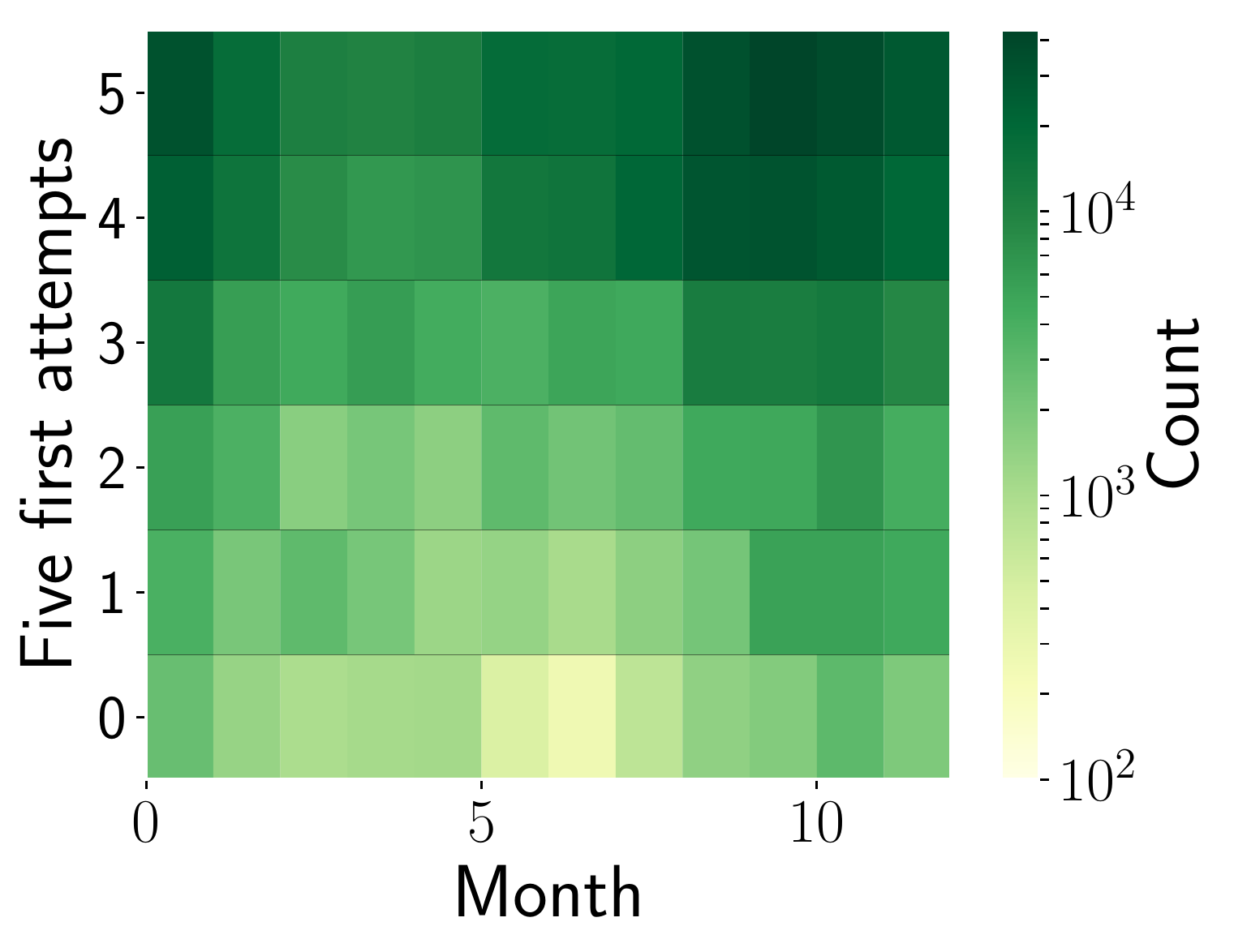} \\
  (a) Disaggregated data & (b) Number of samples \\
  \includegraphics[width=0.45\linewidth]{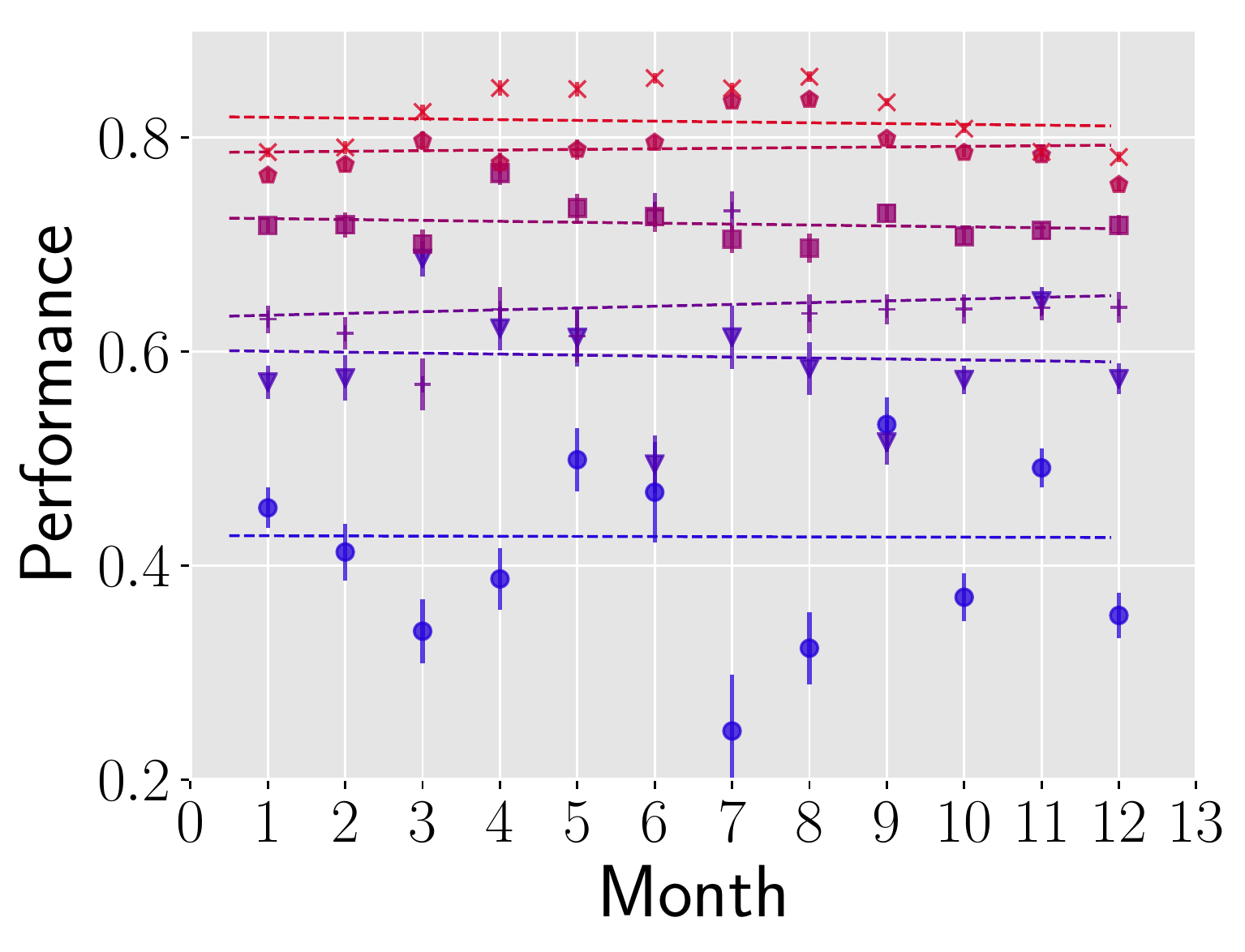} &
  \includegraphics[width=0.45\linewidth]{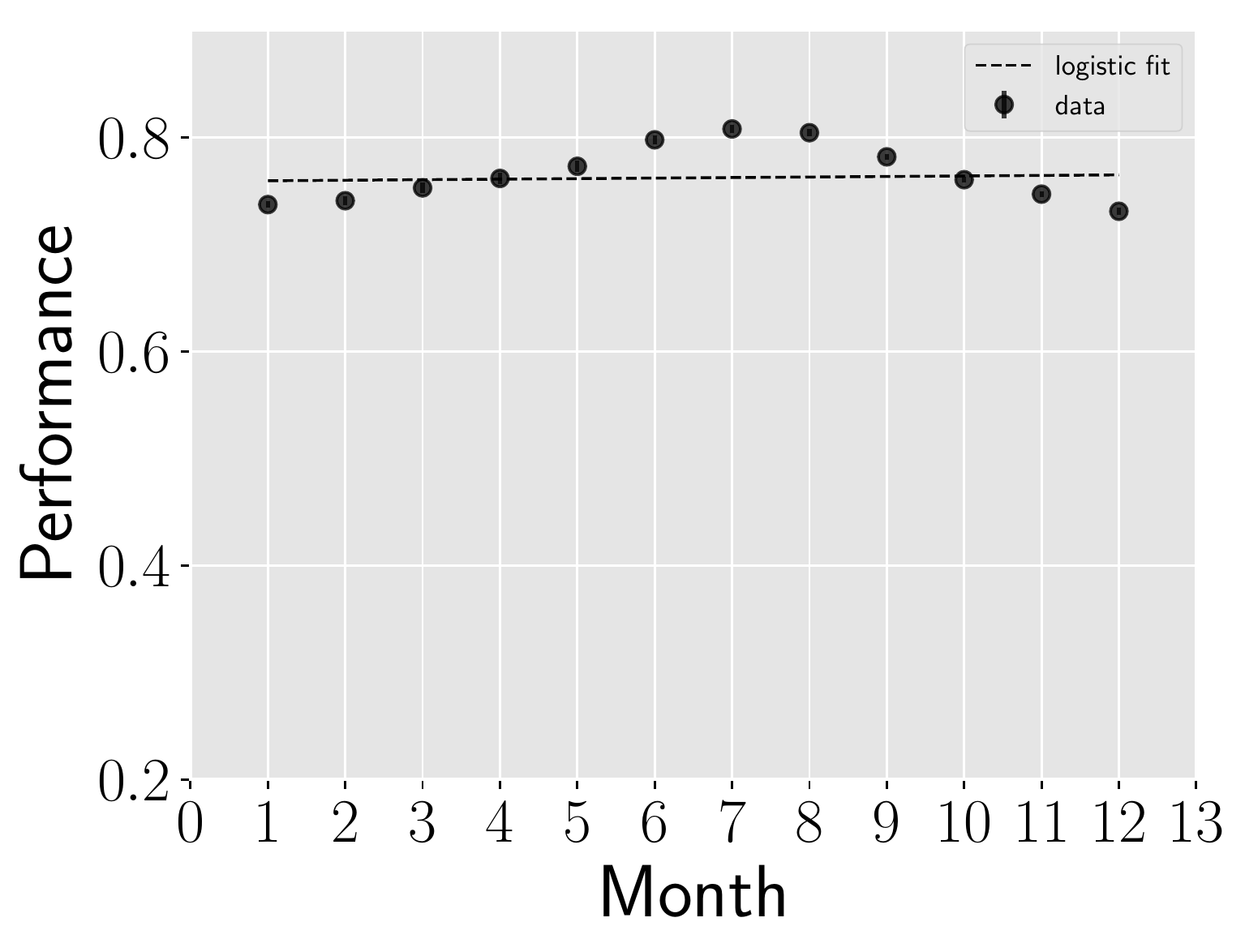}  \\
   (c) Subgroup trends & (d) Aggregate trend
\end{tabular}
\caption{Disaggregation of Khan Academy data showing performance as a function of \emph{month}, conditioned on \emph{five first attempts}. (a) The heat map shows average performance as a function of the \emph{month}.
(b) Number of data samples within each subgroup.
The trends in (c) the disaggregated data and in (d) aggregated data. %Error bars in (c) and (d) show the 95\% confidence interval.
\label{fig:ka_paradox}}
\end{figure}

Figure~\ref{fig:ka_paradox} shows the disaggregation corresponding to the covariate \emph{month}, conditioned on \emph{five first attempts}.
When data is aggregated over the entire population, there appears to be a slight seasonal variation, with performance higher on average during the summer months (Fig.~\ref{fig:ka_paradox}(d)). Once data is disaggregated by \emph{five first attempts}, the seasonal trends are no longer so obvious in several subgroups (Fig.~\ref{fig:ka_paradox}(c)).
Interestingly, it appears to be the high achieving users (who correctly answer more of the five first problems), who perform better during the summer months. This suggests that population of KA changes over the course of the year, with motivated, high achieving students using the platform during their summer break.

\subsection{Duolingo}
Duolingo (DL) is an online language learning platform, which allows users to learn dozens of different languages. DL offers a gamified learning environment, where users progress through levels by practicing vocabulary and dictation skills. The DL halflife-regression~\cite{settles.acl16} dataset (https://github.com/duolingo/halflife-regression) follows a subset of learners over a period of two weeks. Users are shown vocabulary words and asked to recall them correctly. Users may be shown between 7 and 20 words per lesson, and may have multiple lessons in a session. Sessions are defined in a similar way as before---a period of activity without a break longer than one hour.

Users in general perform quite well, correctly recalling a large number of words in a lesson. This makes it difficult to discern changes in performance. Therefore, we define performance in a more stringent way, as a binary variable, which is equal to one if the user had perfect performance (i.e., correctly recalled all words in a lesson), and zero otherwise.  We used more than two dozen features to describe performance. These include the number of words seen and correctly answered  during a lesson (\emph{lesson seen} and \emph{lesson correct}), the number of \emph{distinct words} shown during a lesson, \emph{lesson index} among all lessons for this user, \emph{time to next lesson}, \emph{time since the previous lesson}, \emph{lesson position} within its session, \emph{session length} in terms of the number of lessons and \emph{duration}, etc. User-related features include the number of \emph{five first lessons} correctly answers, number of \emph{all perfect lessons} with perfect performance, total \emph{number of lessons},  the \emph{total number of words seen} and the \emph{correctly} answered, and the \emph{time} the user was active.

\begin{table}
  \caption{Variables defining important disaggregations of Duolingo data, along with their pseudo-$R^2$ scores.}
  \label{tab:dl_pairs}
\begin{center}
\scalebox{0.9}{
  \begin{tabular}{|@{ }c@{ }|@{ }c|@{ }c@{ }|}
    \hline
    \textbf{$R^2_{Mc}$}& \emph{Covariate} \textbf{$X_j$}& \emph{Conditioning on} \textbf{$X_c$}\\% &\textbf{$\mathcal{S}\%$}\\
    \hline
    \hline
0.08 & Lesson position & All perfect lessons \\
\hline
0.11 & Lesson index & All perfect lessons \\
0.09 & Lesson index & First five lessons \\
\hline
0.16 & Number of lessons & All perfect lessons \\
0.09 & Number of lessons & First five lessons \\
\hline
0.11 & Number of sessions & All perfect lessons \\
0.09 & Number of sessions & First five lessons \\
0.05 & Number of sessions & Session seen \\
\hline
0.1 & Session number & All perfect lessons \\
0.09 & Session number & First five lessons \\
0.05 & Session number & Session seen \\
0.05 & Session number & Session correct \\
0.05 & Session number & Distinct words \\
0.02 & Session number & Time since previous lesson \\
\hline
0.09 & Session length & All perfect lessons \\
\hline
0.06 & Session correct & Distinct words \\
\hline
0.09 & Session duration & First five lessons \\
0.09 & Session duration & All perfect lessons \\
0.0 & Session duration & Session length \\
\hline
0.08 & Time since previous lesson & All perfect lessons \\
\hline
\end{tabular}
}
\end{center}
\end{table}

Of the 462 potential disaggregations of DL data, 51 were found to be significant using the $\chi^2$ test.
Table~\ref{tab:dl_pairs} reports disaggregations associated with select covariates, including \emph{lesson's position} within a session, \emph{lesson index} in user's history, the \emph{number of lessons} the user completed, etc. The trends with respect to some of the covariates could be explained by several different disaggregations, with some of them having relatively high values of pseudo-$R^2$. Again, user experience (\emph{all perfect lessons}) and initial skill (\emph{five first lessons}) appear as significant conditioning variables.

\begin{figure}[h!]
\centering
\begin{tabular}{@{}l@{}l@{}}
  \includegraphics[width=0.5\linewidth]{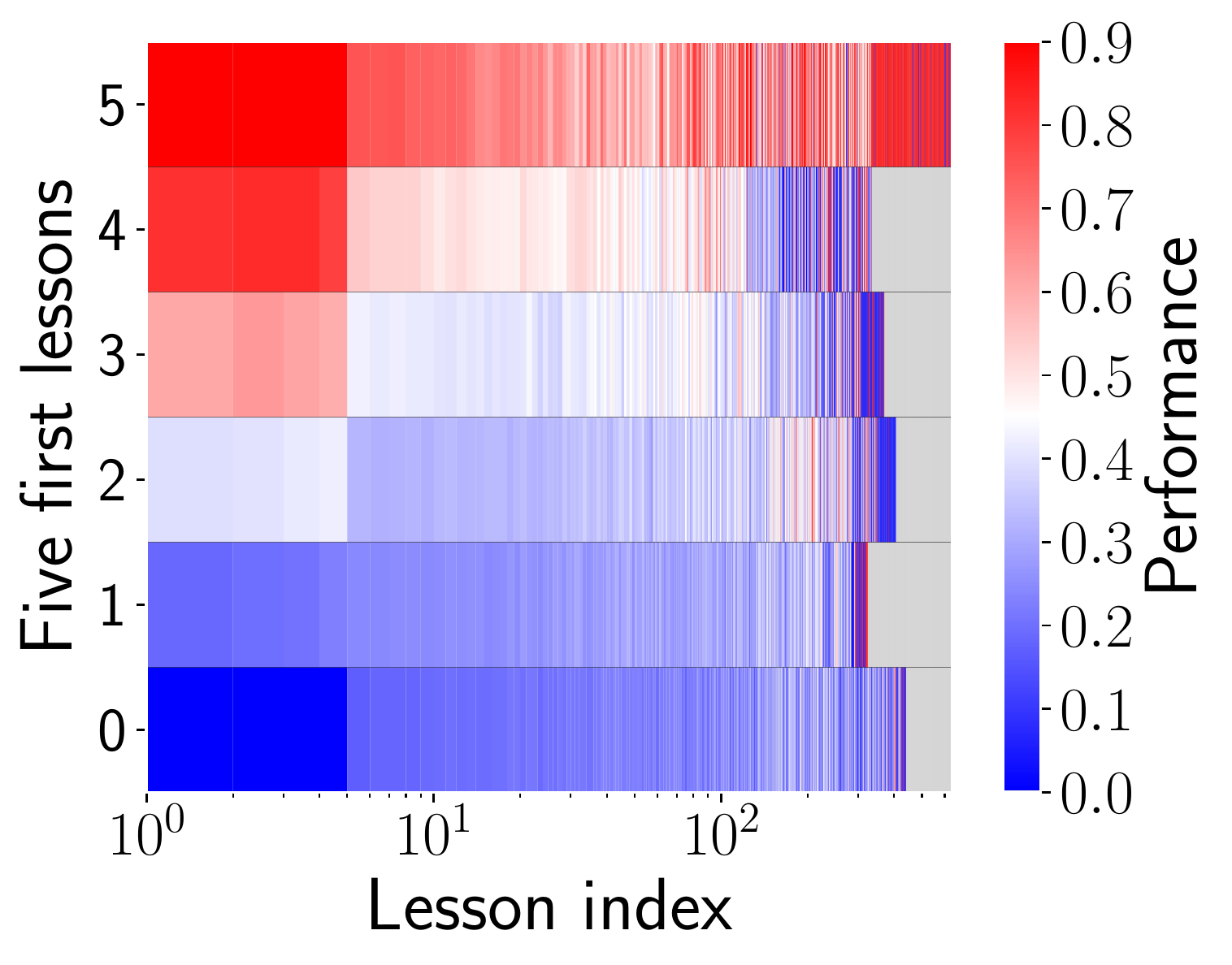} &
  \includegraphics[width=0.5\linewidth]{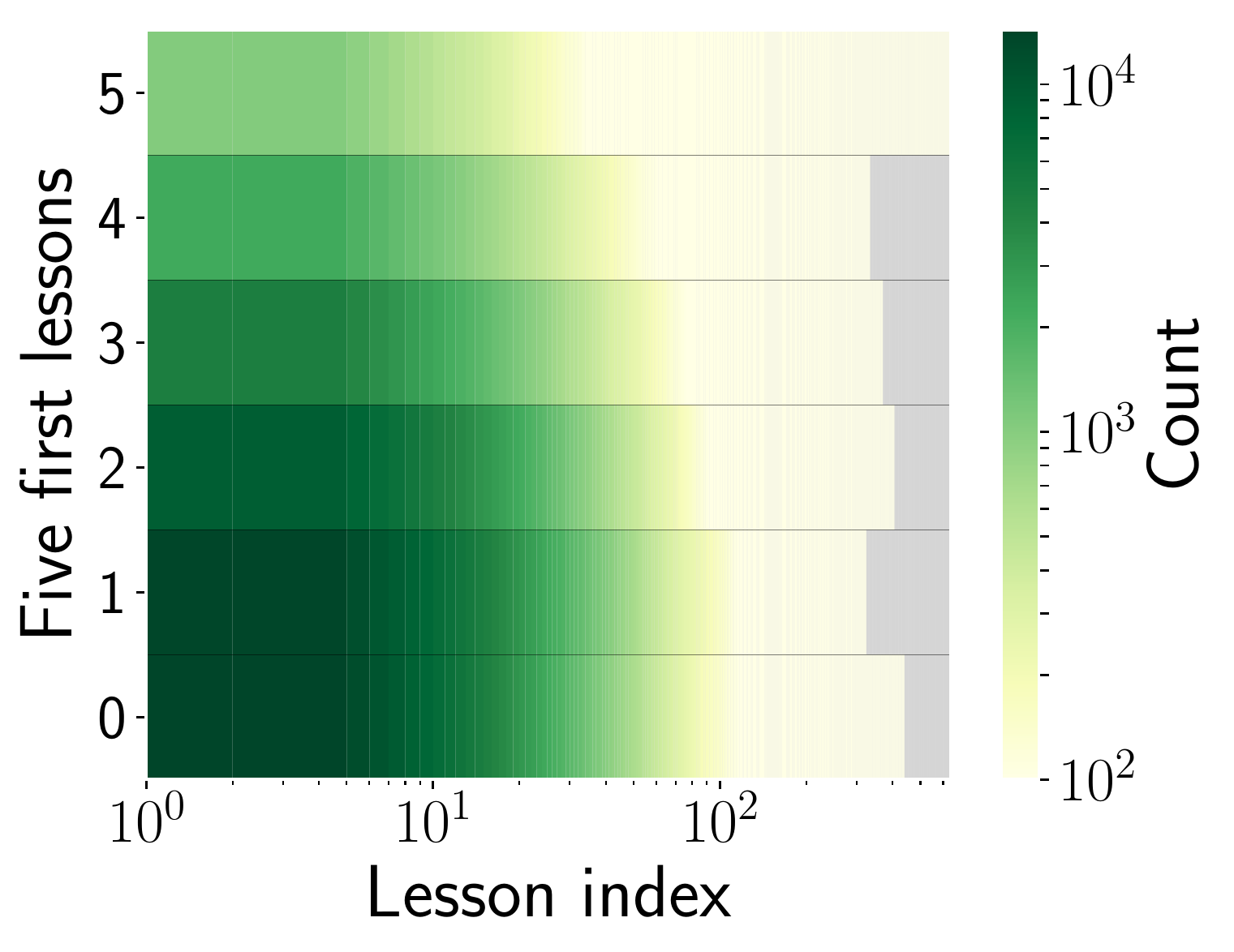} \\
  (a) Disaggregated data & (b) Number of samples \\
  \includegraphics[width=0.45\linewidth]{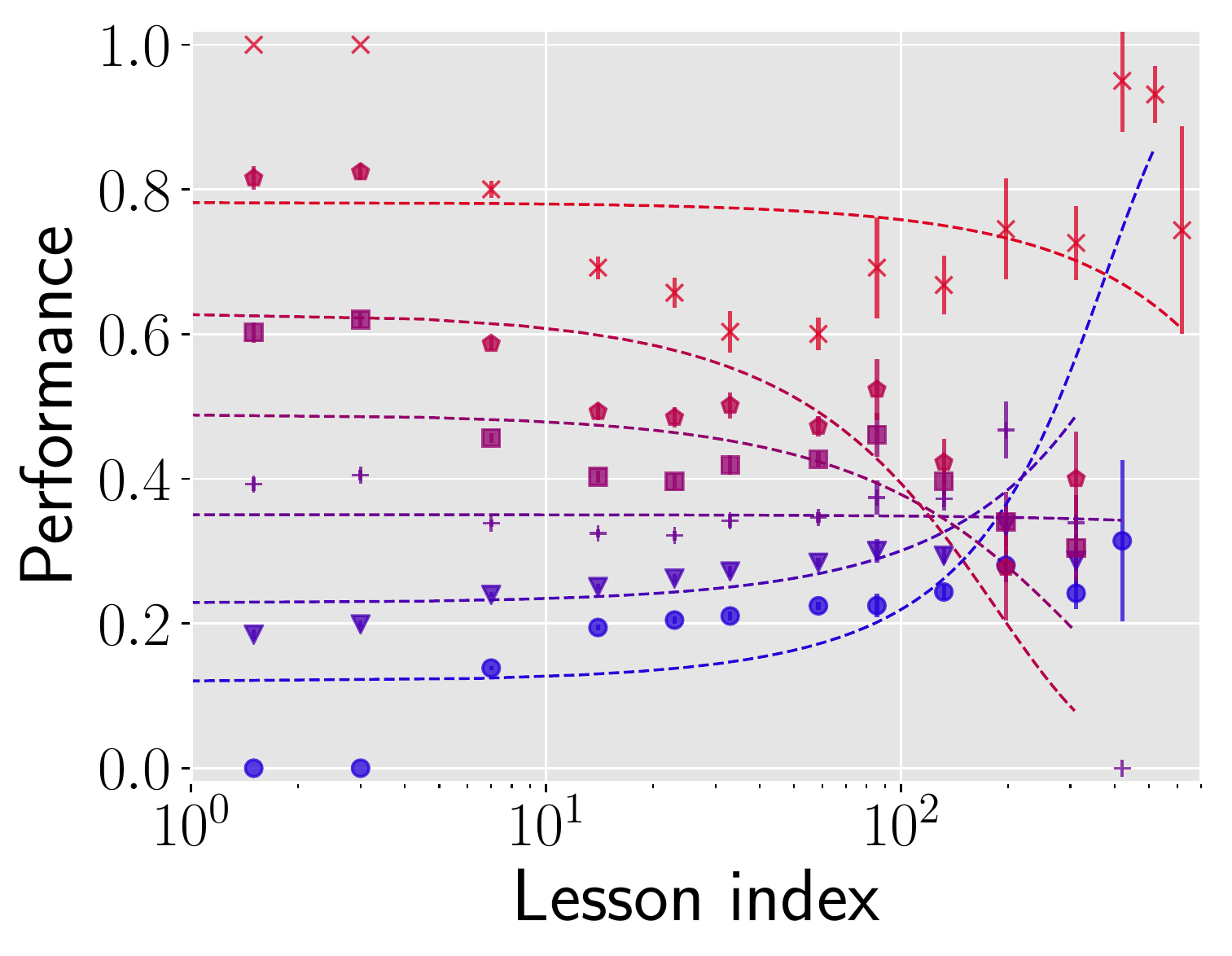}  &
  \includegraphics[width=0.45\linewidth]{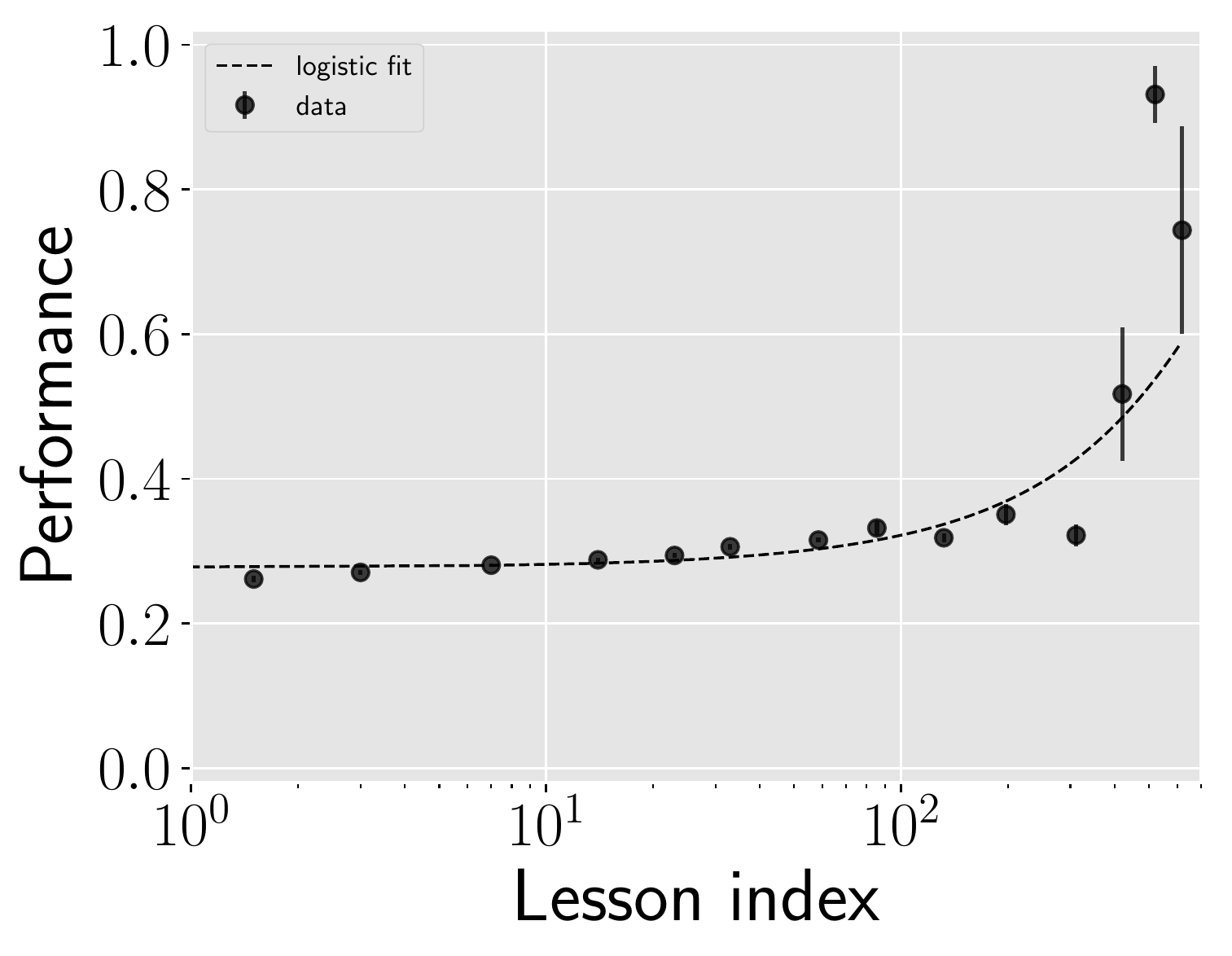} \\
   (c) Subgroup trends & (d) Aggregate trend
\end{tabular}
\caption{Disaggregation of Duolingo data.  (a) The heat map shows performance, %i.e., probability to answer all the words correctly,
as a function of how many lessons the user completed, conditioned on how many of the \emph{five first lessons} were answered correctly. (b) Number of data samples within each bin of the heat map.
Trends in (c) the disaggregated data and in (d) aggregate data. Errors bars show 95\% confidence interval.
\label{fig:dl_paradox}}
\end{figure}

Figure~\ref{fig:dl_paradox} examines the impact of experience on performance. In the aggregate data Fig.~\ref{fig:dl_paradox}(d), performance appears to increase as function of experience (\emph{lesson index}): users who have more practice perform better. However, once the data is disaggregated by initial performance (\emph{five first lessons}), or skill, in Fig.~\ref{fig:dl_paradox}(c), a subtler picture emerges. Users who initially performed the worst (bottom bins in Fig.~\ref{fig:dl_paradox}(a)) improve their performance as they have more lessons, while the best performers initially (top bins) decline. This may be due to ``regression to the mean'', as pure luck could have helped the initially best performers and hurt the initially worst performers.

\begin{figure}[h!]
\centering
\begin{tabular}{@{}l@{}l@{}}
  \includegraphics[width=0.5\linewidth]{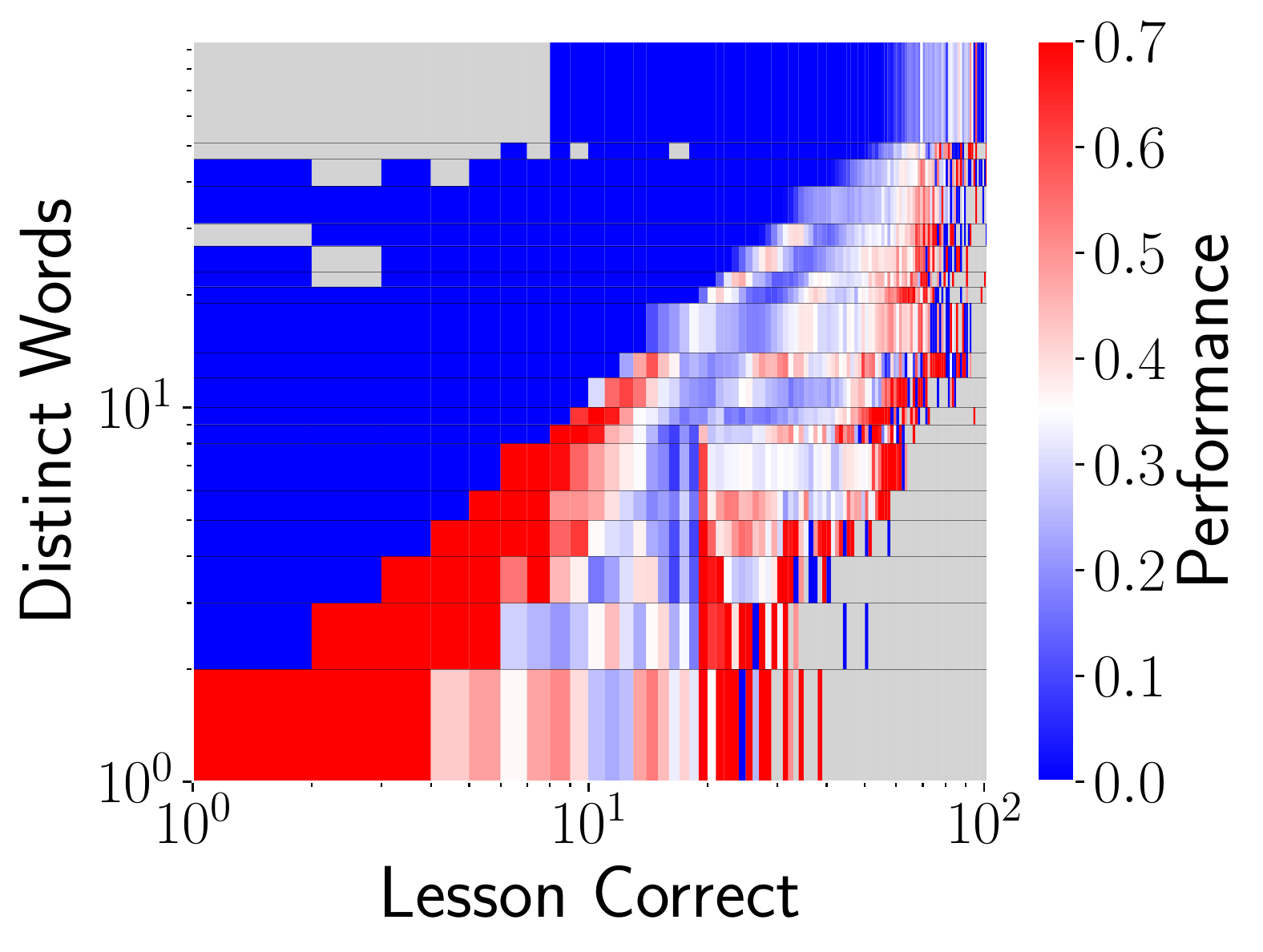} &
  \includegraphics[width=0.5\linewidth]{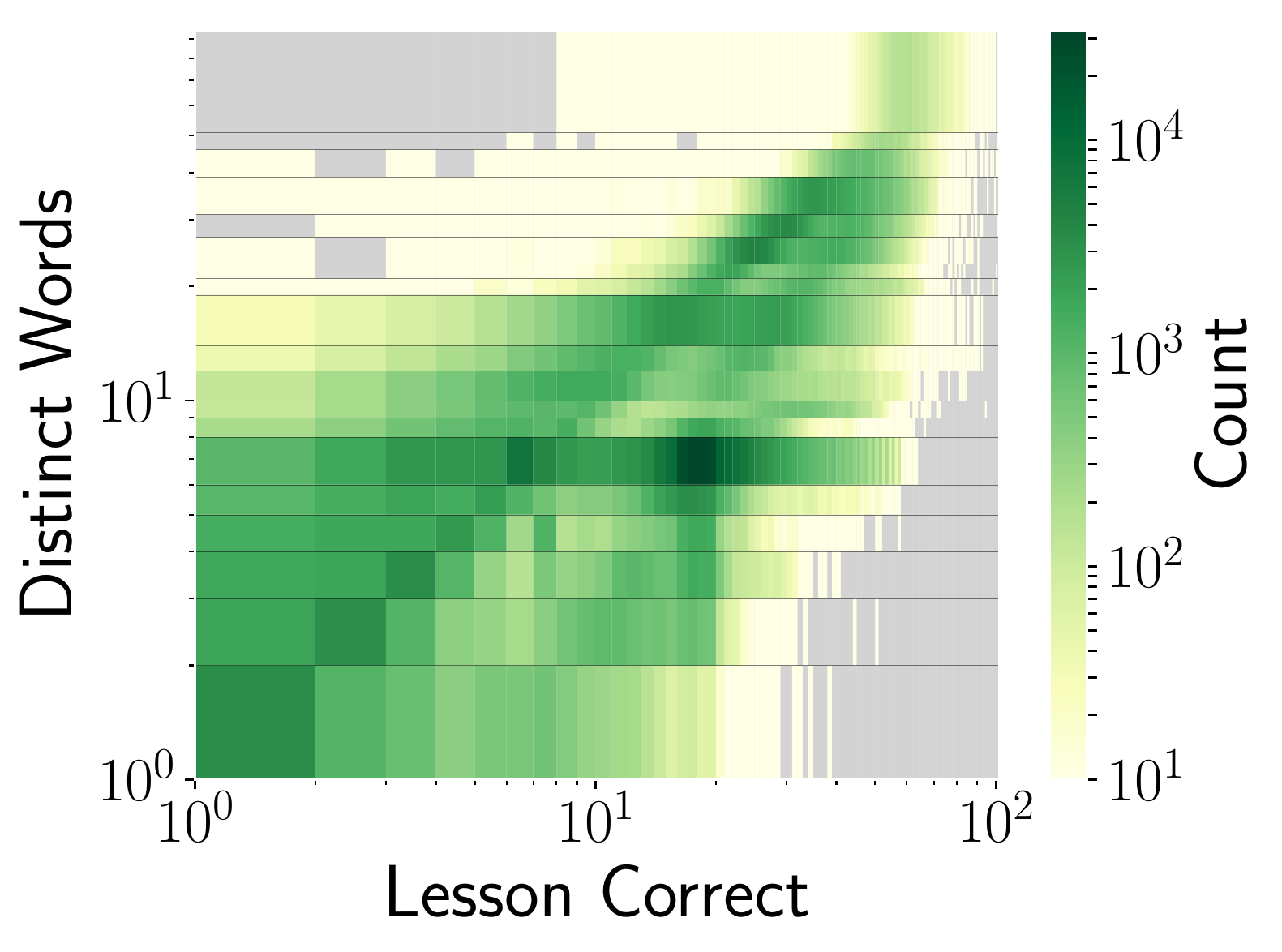} \\
  (a) Disaggregated data & (b) Number of samples \\
  \includegraphics[width=0.45\linewidth]{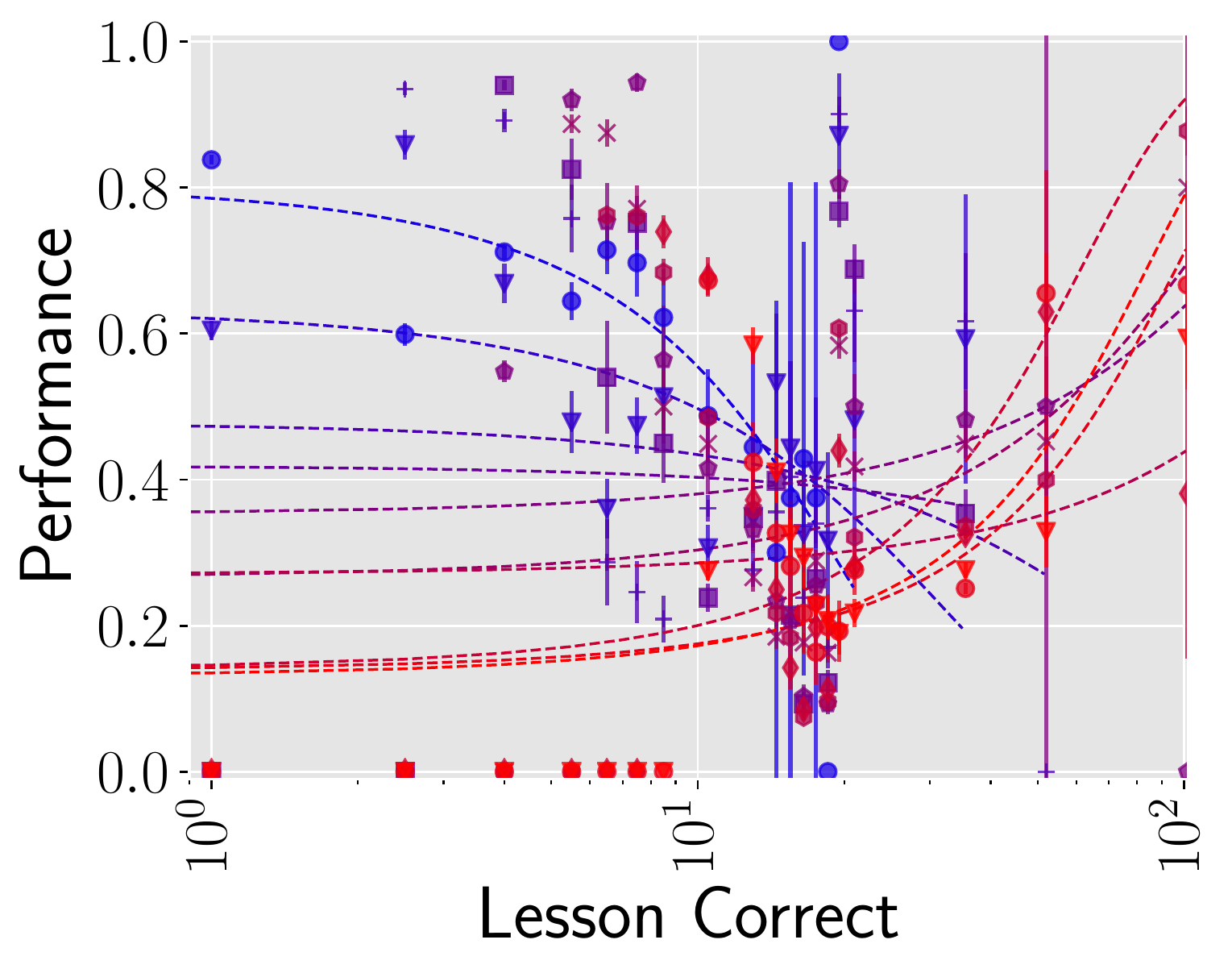}  &
  \includegraphics[width=0.45\linewidth]{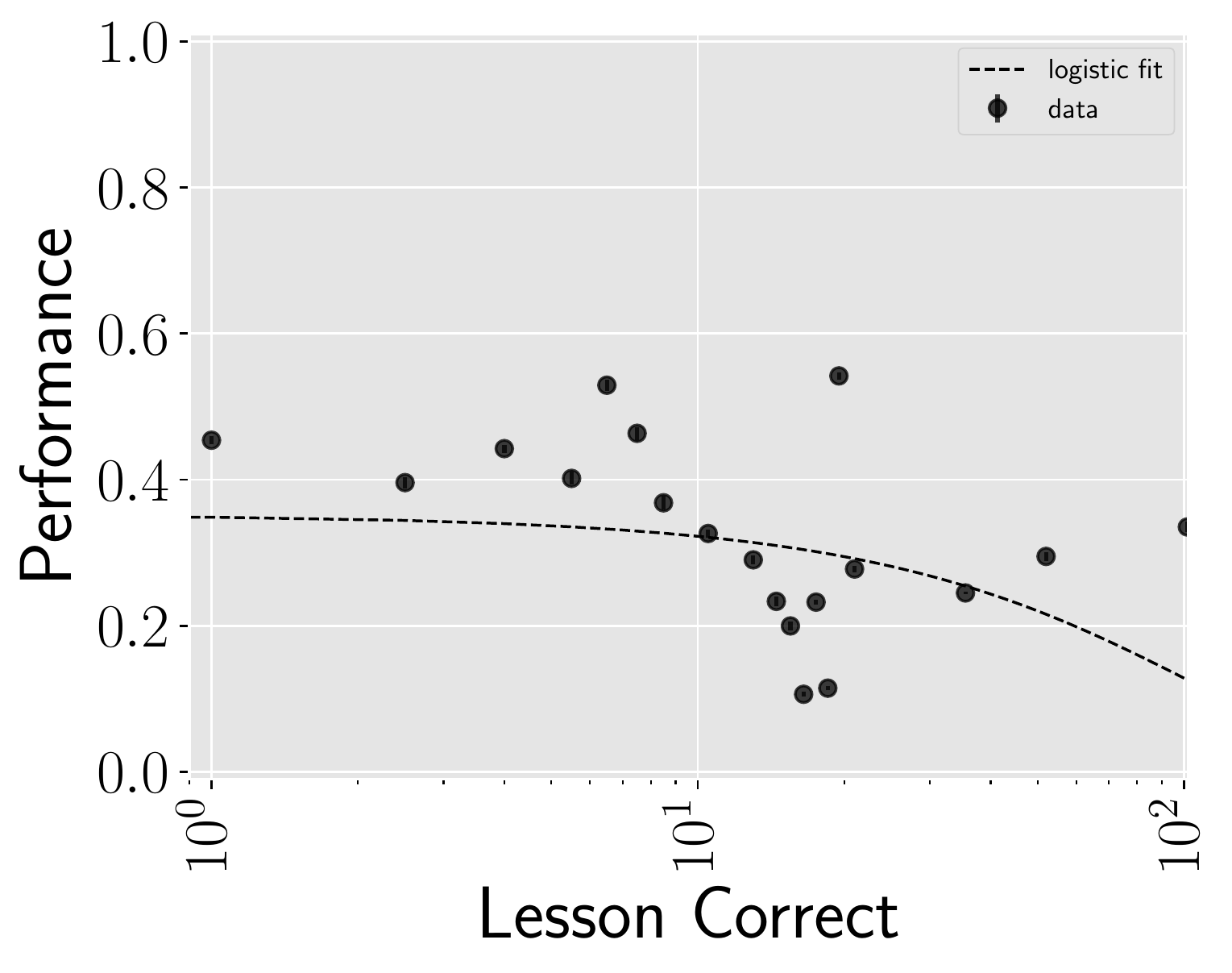} \\
   (c) Subgroup trends & (d) Aggregate trend
\end{tabular}
\caption{Disaggregation of Duolingo data showing performance as a function of \emph{lesson Correct} (a) The heat map shows performance, i.e., probability to answer all the words correctly, conditioned on the number of \emph{distinct words} in the lesson. (b) Number of data samples in each bin of the heat map.
Trends in (c) the disaggregated data and in (d) aggregate data. Errors bars show 95\% confidence interval.
\label{fig:dl2_paradox}}
\end{figure}

Another disaggregation of DL data is shown in Figure~\ref{fig:dl2_paradox}. The plots show performance as a function of \emph{lesson correct}, the number of words correctly answered in a lesson. In the aggregate data, performance shows an overall decline; however, conditioned on \emph{distinct words} (the total number of unique words shown in a lesson), performance shows more complex trends. The red values appearing initially along the diagonal show perfect lessons, where users answered all words they were shown correctly. However, as the lessons become more difficult---more distinct words are introduced---it becomes more difficult for users to have perfect performance. After 20 new words are shown in a lesson, users can no longer answer all the words correctly. Also interesting is a region of lower performance starting around values of \emph{lesson correct} near 20 and \emph{distinct words} between 3 and 10, and continues upwards and to the right. For some reason user performance drops in this regime.

% ______________________________________________________________________
%                                                 CONCLUSION
% ______________________________________________________________________
\subsection{Discussion}
There are several commonalities emerging from the three data sets we studied. Across platforms,  initial performance, captured by \emph{first five attempts} in the KA data or \emph{first five lessons} in the DL data, appeared as an important conditioning variable differentiating the subgroups. Those users who were initially high performers appear to be different from the low performers, especially when looking at how their performance changes over time. While initial performance could capture skill or background knowledge, further analysis is needed to link it to this characteristic.

Experience also appeared as an important feature differentiating users. As a proxy of experience we used such features as the \emph{number of lessons} in DL data, user \emph{tenure} in KA data, and \emph{number of answers} and \emph{reputation} in SE data. However, whether this variable reflects the benefits of practice, or simply captures user motivation, is not clear.

Features linked to user activity differentiated many important subgroups across all three data sets. Features such as \emph{time since previous lesson}, \emph{session length} and \emph{position within a session}, \emph{total time solving} a problem, were all significant conditioning variables. This suggests that performance on these platforms has a non-trivial dynamic component that merits deeper investigation. Indeed, a study of Stack Exchange observed short term deterioration in performance~\cite{Ferrara2017dynamics}. Our study suggests that such an effect may be general.
In addition, temporal features, such as \emph{month}, type of \emph{weekday}, \emph{timestamp}, were found to be important covariates. This indicates that temporal effects can explain differences in performance: e.g., people who use learning web sites on weekends are different from those who use them during the week. Our analysis helps identify such subgroups and understand their behavior.

\section{Conclusion}
We described a method that identifies interesting behaviors within heterogeneous behavioral data by leveraging Simpson's paradox. The method automatically disaggregates data by partitioning it on some conditioning variable, and looks for those conditioning variables that result in trend reversal in many subgroups.
The method ranks these disaggregations based on how well linear models describe the disaggregated data compared to how well they describe population as a whole. These disaggregated subgroups are interesting because their behavior is significantly different from that of the remainder of the population, which implies that important behavioral differences exist within the population.

We illustrated the use of the method as a data exploration tool by applying it to study human performance on three online platforms, including question-answering site Stack Exchange, and online learning sites Khan Academy and Duolingo. Our method identified skill (judged from user's initial performance on the site) and experience (how long the user has been active on the site) as important features differentiating user performance.

Conditioning on a variable to make subgroups more homogenous is the important and the first step in our method; however, subgroups may still be heterogeneous. As a future direction, we can use multiple features for conditioning on the data to make the subgroups even more homogenous, afterward we can look at an independent variable trend reversal in these subgroups.
We have used our method for binary outcome variables, however there are also continues outcome variables in behavioral data. Our method can be extended to more general forms, like GLM, to support all types of outcome variable. However, new trend analysis algorithm needs different statistical methods as a goodness of fit measure.
In addition, preliminary explorations suggest that pairs of variables with high $R^2_{McFadden}$ value could be used in combination to make a new variable which is highly correlated with outcome variable. For example, in KA data, pair (\emph{all first attempts}, \emph{all attempts}) has the highest value of pseudo-$R^2$ among all pairs. We can define new variable as ratio of \emph{number of correctly solved question on the first attempt} to characterize user performance during his or her tenure. This new variable is highly correlated with outcome variable, \emph{performance}. The same thing is happened for pairs
(\emph{session seen}, \emph{session correct}) in Duolingo, and (\emph{number of answers}, \emph{Reputation}) in Stack Exchange.

\section{Acknowledgments}
This material is based upon work supported by the U.S. Army Small Business Innovation Research Program Office, Defense Advanced Research Projects Agency (DARPA) and the Army Research Office under Contract No. W911NF-18-C-0011, and the James S. McDonnell Foundation.

\small
%\bibliographystyle{aaai}
%\bibliography{references}

\begin{thebibliography}{}

\bibitem[\protect\citeauthoryear{Alipourfard, Fennell, and
  Lerman}{2018}]{Alipourfard2018}
Alipourfard, N.; Fennell, P.~G.; and Lerman, K.
\newblock 2018.
\newblock Can you trust the trend: Discovering simpson's paradoxes in social
  data.
\newblock {\em arXiv preprint arXiv:1801.04385}.

\bibitem[\protect\citeauthoryear{Bickel, Hammel, and
  O'Connell}{1975}]{Bickel1975}
Bickel, P.~J.; Hammel, E.~A.; and O'Connell, J.~W.
\newblock 1975.
\newblock Sex bias in graduate admissions: Data from berkeley.
\newblock {\em Science} 187(4175):398--404.

\bibitem[\protect\citeauthoryear{Blyth}{1972}]{Blyth1972}
Blyth, C.~R.
\newblock 1972.
\newblock On simpson's paradox and the sure-thing principle.
\newblock {\em Journal of the American Statistical Association}
  67(338):364--366.

\bibitem[\protect\citeauthoryear{Bond \bgroup et al\mbox.\egroup
  }{2012}]{bond201261}
Bond, R.~M.; Fariss, C.~J.; Jones, J.~J.; Kramer, A.~D.; Marlow, C.; Settle,
  J.~E.; and Fowler, J.~H.
\newblock 2012.
\newblock A 61-million-person experiment in social influence and political
  mobilization.
\newblock {\em Nature} 489(7415):295--298.

\bibitem[\protect\citeauthoryear{Fabris and
  Freitas}{2000}]{fabris2000discovering}
Fabris, C., and Freitas, A.
\newblock 2000.
\newblock Discovering surprising patterns by detecting occurrences of simpson's
  paradox.
\newblock In Bramer, M.; Macintosh, A.; and Coenen, F., eds., {\em Research and
  Development in Intelligent Systems XVI}. Springer London.
\newblock  148--160.

\bibitem[\protect\citeauthoryear{Ferrara \bgroup et al\mbox.\egroup
  }{2017}]{Ferrara2017dynamics}
Ferrara, E.; Alipoufard, N.; Burghardt, K.; Gopal, C.; and Lerman, K.
\newblock 2017.
\newblock Dynamics of content quality in collaborative knowledge production.
\newblock In {\em Proceedings of 11th AAAI International Conference on Web and
  Social Media}.
\newblock AAAI.

\bibitem[\protect\citeauthoryear{Hosmer~Jr, Lemeshow, and
  Sturdivant}{2013}]{deviance}
Hosmer~Jr, D.~W.; Lemeshow, S.; and Sturdivant, R.~X.
\newblock 2013.
\newblock Applied logistic regression.
\newblock volume 398.
\newblock John Wiley \& Sons.

\bibitem[\protect\citeauthoryear{Kincaid \bgroup et al\mbox.\egroup
  }{1975}]{Readability}
Kincaid, J.~P.; R.~P.~Fishburnea, J.; Rogers, R.~L.; and Chissom, B.~S.
\newblock 1975.
\newblock Derivation of new readability formulas (automated readability index,
  fog count and flesch reading ease formula) for navy enlisted personnel.
\newblock Technical report, U.S. Navy.

\bibitem[\protect\citeauthoryear{Kleinberg \bgroup et al\mbox.\egroup
  }{2017}]{kleinberg2017human}
Kleinberg, J.; Lakkaraju, H.; Leskovec, J.; Ludwig, J.; and Mullainathan, S.
\newblock 2017.
\newblock Human decisions and machine predictions.
\newblock Technical report, National Bureau of Economic Research.

\bibitem[\protect\citeauthoryear{Lazer \bgroup et al\mbox.\egroup
  }{2009}]{Lazer09}
Lazer, D.; Pentland, A.; Adamic, L.; Aral, S.; Barab\'{a}si, A.-L.; Brewer, D.;
  Christakis, N.; Contractor, N.; Fowler, J.; Gutmann, M.; and Jebara, T.
\newblock 2009.
\newblock Computational social science.
\newblock {\em Science} 323:721--723.

\bibitem[\protect\citeauthoryear{Lerman}{2018}]{Lerman2018jcss}
Lerman, K.
\newblock 2018.
\newblock Computational social scientist beware: Simpson's paradox in
  behavioral data.
\newblock {\em Journal of Computational Social Science} 1(1):49--58.

\bibitem[\protect\citeauthoryear{McFadden and others}{1973}]{mcfadden1}
McFadden, D., et~al.
\newblock 1973.
\newblock In {\em Conditional logit analysis of qualitative choice behavior},
  121.
\newblock Institute of Urban and Regional Development, University of
  California.

\bibitem[\protect\citeauthoryear{McFadden and others}{1977}]{mcfadden2}
McFadden, D., et~al.
\newblock 1977.
\newblock In {\em Quantitative methods for analyzing travel behavior of
  individuals: some recent developments},  307.
\newblock Institute of Transportation Studies, University of California.

\bibitem[\protect\citeauthoryear{McFarland, Lewis, and
  Goldberg}{2016}]{mcfarland2016sociology}
McFarland, D.~A.; Lewis, K.; and Goldberg, A.
\newblock 2016.
\newblock Sociology in the era of big data: The ascent of forensic social
  science.
\newblock {\em The American Sociologist} 47(1):12--35.

\bibitem[\protect\citeauthoryear{Norton and Divine}{2015}]{Norton2015simpson}
Norton, H.~J., and Divine, G.
\newblock 2015.
\newblock Simpson's paradox … and how to avoid it.
\newblock {\em Significance} 12(4):40--43.

\bibitem[\protect\citeauthoryear{Settles and Meeder}{2016}]{settles.acl16}
Settles, B., and Meeder, B.
\newblock 2016.
\newblock A trainable spaced repetition model for language learning.
\newblock In {\em Proceedings of the Association for Computational Linguistics
  (ACL)},  1848--1858.
\newblock ACL.

\bibitem[\protect\citeauthoryear{Vaupel and
  Yashin}{1985}]{Vaupel85heterogeneity}
Vaupel, J.~W., and Yashin, A.~I.
\newblock 1985.
\newblock Heterogeneity's ruses: some surprising effects of selection on
  population dynamics.
\newblock {\em The American Statistician} 39(3):176--185.

\end{thebibliography}

\end{document}